\documentclass[aps,prb,preprint,superscriptaddress,reprint,preprintnumbers,amssymb,longbibliography]{revtex4-1}
\pdfoutput=1

\usepackage{amsmath}
\usepackage{braket}
\usepackage{bbold}
\usepackage[version=3]{mhchem}
\usepackage{graphicx}
\usepackage{amsmath}
\usepackage{color}
\usepackage{xcolor}
\usepackage{units}
 \usepackage[caption=false]{subfig}
\usepackage{multirow}
\usepackage{hyperref}

\DeclareMathOperator{\Tr}{Tr}
\DeclareMathOperator{\erf}{erf}

\AtBeginDocument{\usepackage{booktabs}}

\newcommand{\SM}[1]{\textcolor{red}{[SM: #1]}}

 \definecolor{rankingcolor1}{RGB}{ 77,175, 74} 
 \definecolor{rankingcolor2}{RGB}{ 55,126,184} 
 \definecolor{rankingcolor3}{RGB}{228, 26, 28} 

\begin{document}

\title{\SM{title to be discussed}\textsc{CheSS}: A library for efficient linear scaling electronic structure calculations using sparse Chebyshev matrix polynomials}
\title{Efficient computation of sparse matrix functions for large scale electronic structure calculations: The \textsc{CheSS} library}


\author{Stephan Mohr}
\affiliation{Barcelona Supercomputing Center (BSC)}
\email{stephan.mohr@bsc.es}

\author{William Dawson}
\affiliation{RIKEN Advanced Institute for Computational Science, Kobe, Japan, 650-0002}

\author{Michael Wagner}
\affiliation{Barcelona Supercomputing Center (BSC)}

\author{Damien Caliste}
\affiliation{Univ.\ Grenoble Alpes, INAC-MEM, L\_Sim, F-38000 Grenoble, France}
\affiliation{CEA, INAC-MEM, L\_Sim, F-38000 Grenoble, France}

\author{Takahito Nakajima}
\affiliation{RIKEN Advanced Institute for Computational Science, Kobe, Japan, 650-0002}

\author{Luigi Genovese}
\affiliation{Univ.\ Grenoble Alpes, INAC-MEM, L\_Sim, F-38000 Grenoble, France}
\affiliation{CEA, INAC-MEM, L\_Sim, F-38000 Grenoble, France}


\date{\today}

\begin{abstract}
We present \textsc{CheSS}, the ``Chebyshev Sparse Solvers'' library, which has been designed to solve typical problems arising in large scale electronic structure calculations
using localized basis sets.
The library is based on a flexible and efficient expansion in terms of Chebyshev polynomials, 
and presently features the calculation of the density matrix, the calculation of matrix powers for arbitrary powers, 
and the extraction of eigenvalues in a selected interval.
\textsc{CheSS} is able to exploit the sparsity of the matrices and scales linearly with respect to the number of non-zero entries, making it well suited for large scale calculations.
The approach is particularly adapted for setups leading to small spectral widths of the involved matrices, and outperforms alternative methods in this regime.
By coupling \textsc{CheSS} to the DFT code \textsc{BigDFT} we show that such a favorable setup is indeed possible in practice.
In addition, the approach based on Chebyshev polynomials can be massively parallelized, and CheSS exhibits excellent scaling up to thousands of cores even for relatively small matrix sizes.
\end{abstract}

\maketitle

\section{Introduction}
Sparse matrices are abundant in many branches of science, be it due to the characteristics of the employed basis set (e.g.\ finite elements, wavelets, Gaussians, etc.) or due to intrinsic localization properties of the system under investigation.
Ideally, an operator acting on such matrices should exploit this sparsity as much as possible in order to reach a high efficiency.
Due to the great variety of specific needs there is no simple and unique approach to perform this general task, and various solutions have been conceived to satisfy the respective demands~\cite{higham-functions-2008}.

The \textsc{CheSS} library, which we present in this paper, has its origins in electronic structure calculations, in particular Density Functional Theory (DFT)~\cite{hohenberg-inhomogeneous-1964,kohn-self_consistent-1965}, and is consequently capable of performing the specific matrix operations required in this context.
In principle, these can all be solved straightforwardly using highly optimized Linear Algebra libraries such as \textsc{LAPACK}~\cite{anderson-lapack-1999}/\textsc{ScaLAPACK}\cite{blackford-scalapack-1997}, which is indeed the fastest solution for small to medium size matrices.
However, this approach gets increasingly expensive for large systems, since it uses the matrices in their dense form, which exhibits an inherent cubic scaling.
With the recent widespread availability of DFT codes that are able to tackle the regimes of many thousands atoms systems, such challenging large scale calculations are becoming more and more abundant~\cite{ratcliff-2017-challenges}, and the above mentioned cubic scaling is clearly a serious limitation.

An overview of popular electronic structure codes, in particular focusing on large scale calculations, can be found in Ref.~\citenum{ratcliff-2017-challenges}.
In general these large scale DFT codes work with localized basis sets,
e.g.\ BigDFT~\cite{genovese-daubechies-2008,mohr-2014-daubechies,mohr-2015-accurate}, SIESTA~\cite{soler-the_siesta-2002,artacho-the_siesta-2008}, \textsc{Quickstep}~\cite{vandevondele-quickstep-2005}, ONETEP~\cite{skylaris-introducing-2005,haynes-onetep-2006,mostofi-onetep-2007,skylaris-recent-2008} or Conquest~\cite{bowler-practical-2000,bowler-recent-2006,bowler-an-overview-2010},
and the matrices expressed in these bases 
consequently exhibit a natural sparsity, i.e.\ only those matrix elements where the basis functions overlap are different from zero.
Hence it is advantageous to use --- beyond a certain crossover point --- sparse algorithms instead of the basic dense approaches.
Since exploiting the sparsity is also the key towards linear scaling algorithms --- i.e.\ methods where the computational demand only increases linearly with respect to the system size --- efficient sparse solvers are crucial in this domain. 

The central task within DFT is the calculation of the density matrix $\hat{F}$,
and many different approximate ways of calculating it with a scaling being more favorable than the default cubic one have been derived.
The so-called Fermi Operator Expansion (FOE)~\cite{goedecker-1994-efficient,goedecker-1995-tight-binding}, which gave the inspiration to the creation of \textsc{CheSS}, calculates the density matrix as a direct expansion of the Hamiltonian matrix in terms of Chebyshev polynomials.
Another method, which is similar in spirit, writes the density matrix as a rational expansion~\cite{goedecker-low-1995,goedecker-linear-1999},
exploiting Cauchy's integral theorem in the complex plane. 
This method has the advantage that --- unlike FOE --- it only has to cope with the occupied states, which makes it advantageous for large basis sets containing many high energetic virtual states.
One particular implementation is the PEXSI package~\cite{lin-2013-accelerating},
which we will use later on for a comparison with \textsc{CheSS}.
Other popular approaches to calculate the density matrix are the density-matrix minimization approach~\cite{li-density-matrix-1993}, which calculates the density matrix by minimizing a target function
with respect to $\hat{F}$,
and the divide-and-conquer method~\cite{yang-a-density-1995}, which is based on a partitioning of the density matrix into small subblocks.
An overview over further popular methods --- in particular focusing on linear scaling approaches --- can be found in Refs.~\citenum{goedecker-linear-1999} and \citenum{bowler-O(N)-2012}.

Several studies have compared the various methods~\cite{bates-comparison-1998,daniels-what-1999,liang-improved-2003,jordan-comparison-2005,haynes-density-2008,rudberg-assessment-2011}, but it is very hard --- if not impossible --- to determine the ultimate method that performs best under all circumstances.
Rather all of these methods have their ``niche'' in which they are particularly efficient, and the task of choosing the appropriate approach is therefore strongly influenced by the specific application.
In particular, the choice depends on the physical properties of the system (e.g.\ insulator versus metal), the chosen formalism (e.g.\ all-electron versus pseudopotential), and on the basis set that is used.
These factors influence the properties of the matrices that shall be processed in various ways;
we will focus in the following on the spectral width.
For some of the aforementioned methods this property has only little influence, whereas for others it is crucial.

The other main operation available within \textsc{CheSS}, namely the calculation of a matrix power, where the power can have any --- in particular also non-integer --- value,
is a more general problem appearing also outside of electronic structure calculations.
Among the various powers, calculating the inverse is presumably the most important one, and there exist various approaches to do this efficiently for sparse matrices.
Some approaches allow to calculate the inverse in an efficient way for some special cases.
For banded matrices, for instance, Ran and Huang developed an algorithm~\cite{ran-2009-an_inversion} that is about twice as fast as the standard method based of the $LU$-decomposition.
To determine the diagonal entries of the inverse, Tang and Saad developed a probing method~\cite{tang-2012-a_probing} for situations where the inverse exhibits decay properties.
Another algorithm for the calculation of the diagonal entries is the one by Lin et al.~\cite{lin-2009-fast}, which calculates the diagonal elements of the inverse by hierarchical decomposition of the computational domain.
To get an approximation of the diagonal entries of a matrix, Bekas et al.\ proposed a stochastic estimator~\cite{bekas-2007-an_estimator}.
The \textsc{FIND} algorithm by Li et al.~\cite{li-2008-computing},
which follows the idea of nested dissection~\cite{george-nested-1973}, was as well designed to calculate the exact diagonal entries of Greens functions, but can be extended to calculate any subset of entries of the inverse of a sparse matrix.
In the same way, the Selected Inversion developed by Lin et al.~\cite{lin-a_fast-2011,lin-2011-selinv}, which is based on an $LDL^T$-factorization, allows to exactly calculate selected elements of the inverse.
As for the calculation of the density matrix, the choice of the method for the calculation of matrix powers depends as well on the specific application, and therefore there is no universal best option.
However, we would like to point out again the importance of the spectral width, which might favor or not a particular approach.

In this paper we present with \textsc{CheSS} the implementation of a general approach that can efficiently evaluate a matrix function $f(\mathbf{M})$ using an expansion in Chebyshev polynomials.
Since \textsc{CheSS} has first been used for electronic structure calculations, the functions that are implemented so far are those needed in this context; more details will be given later.
However, there are no restrictions to go beyond this, as the function $f$ can in principle be chosen arbitrarily, with the only restriction that it must be well representable by Chebyshev polynomials over the entire eigenvalue spectrum of $\mathbf{M}$.
\textsc{CheSS} has been particularly designed for matrices with a small eigenvalue spectrum, 
and in this regime it is able to outperform other comparable approaches.
We will show later that such a favorable regime can indeed be reached 
within the context of DFT calculations.

In addition \textsc{CheSS} exploits the sparsity of the involved matrices and hence only calculates those elements that are non-zero.
Obviously, this requires that the solution $f(\mathbf{M})$ can reasonably well be represented within the predefined sparsity pattern;
however, since the sparsity pattern is defined by the underlying physical or mathematical problem, we leave the responsibility of well defining this pattern to the code interfacing \textsc{CheSS}.
If the cost of calculating one matrix element can be considered as constant (which is the case for a high degree of sparsity), we consequently reach an approach that scales linearly with respect to the number of non-zero elements.
Hence, \textsc{CheSS} is an ideal library for linear scaling calculations, which are crucial for the treatment of large systems.

In summary, we present with \textsc{CheSS} a flexible and powerful framework to compute sparse matrix functions required in the context of large scale electronic structure calculations.
If the above mentioned requirement --- namely a small eigenvalue spectrum of the matrices --- is fulfilled, \textsc{CheSS} can yield considerable performance boosts compared to other similar approaches.
Hence, this library represents an interesting tool for any code working with localized basis functions that aims to perform such large scale calculations. 

The remainder of the paper is structured as follows: In Sec.~\ref{sec: Theory} we first show the basic theory behind \textsc{CheSS},
starting with the applicability of \textsc{CheSS} for electronic structure calculations (Sec.~\ref{sec: Applicability of CheSS for electronic structure calculation}),
detailing the basic algorithm (Sec.~\ref{sec: Algorithm}) and available operations (Sec.~\ref{sec: Available operations}), giving a brief discussion about sparsity (Sec.~\ref{sec: Sparsity and truncation}), and finishing with a presentation of our format to store sparse matrices (Sec.~\ref{sec: Storage format of the sparse matrices}).
In Sec.~\ref{sec: Performance} we then give various performance numbers of \textsc{CheSS}, showing the accuracy (Sec.~\ref{sec: Accuracy}), the scaling with matrix properties (Sec.~\ref{sec: Scaling with system properties}), the parallel scaling (Sec.~\ref{sec: Parallel scaling}), and a comparison with other methods (Sec.~\ref{sec: Comparison with other methods}).
Finally we conclude and summarize our work in Sec.~\ref{sec: Conclusions and outlook} and give an outlook on future research.

\section{Motivation and theory}
\label{sec: Theory}

\subsection{Applicability of \textsc{CheSS} for electronic structure calculations}
\label{sec: Applicability of CheSS for electronic structure calculation}
\begin{figure*}
 \subfloat[][solvated DNA\newline(15613 atoms)]{
  \includegraphics[width=0.22\textwidth]{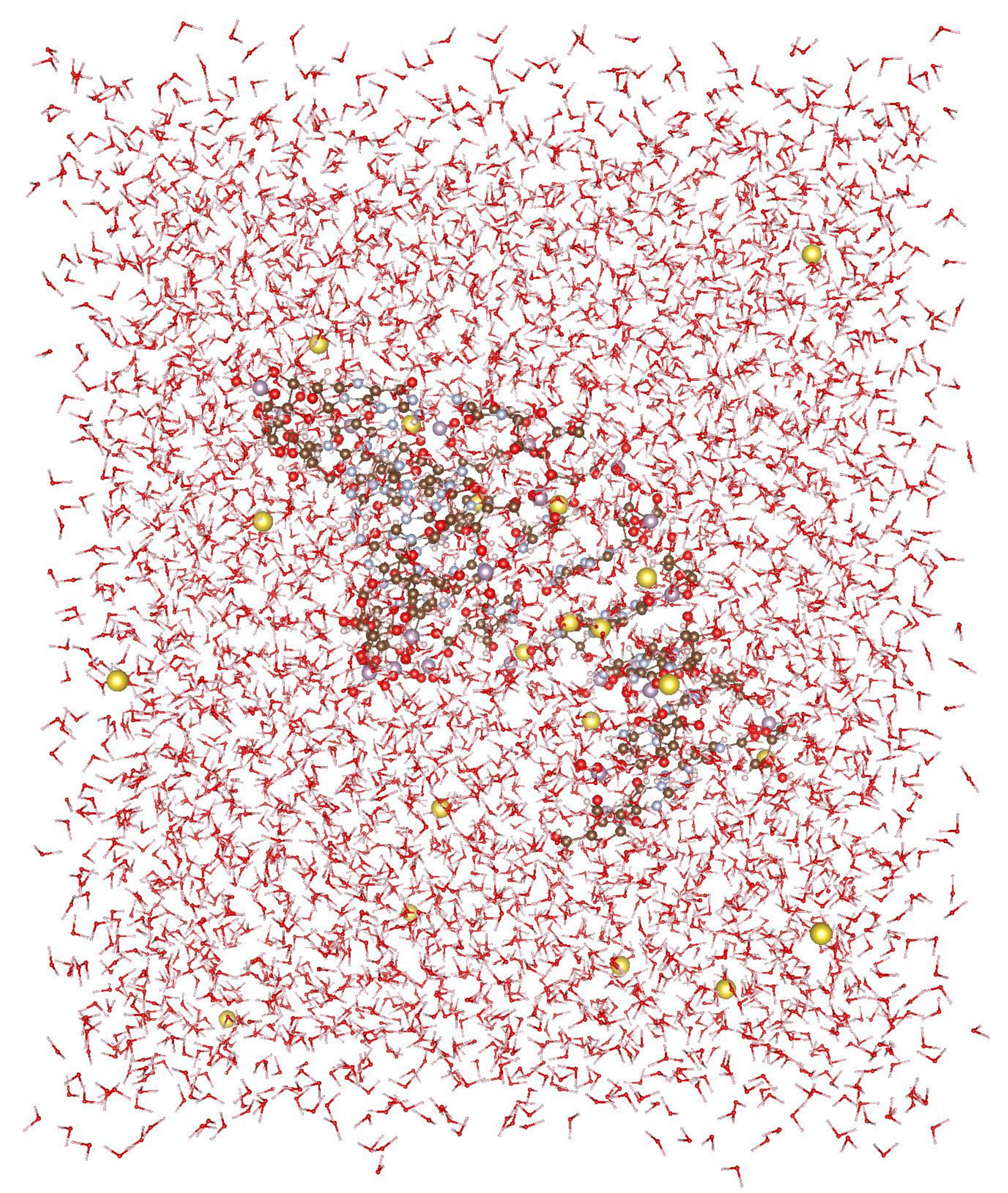}
  \label{fig:DNA}}
  \hspace{10pt}
 \subfloat[][bulk pentacene\newline(6876 atoms)]{
  \includegraphics[width=0.21\textwidth]{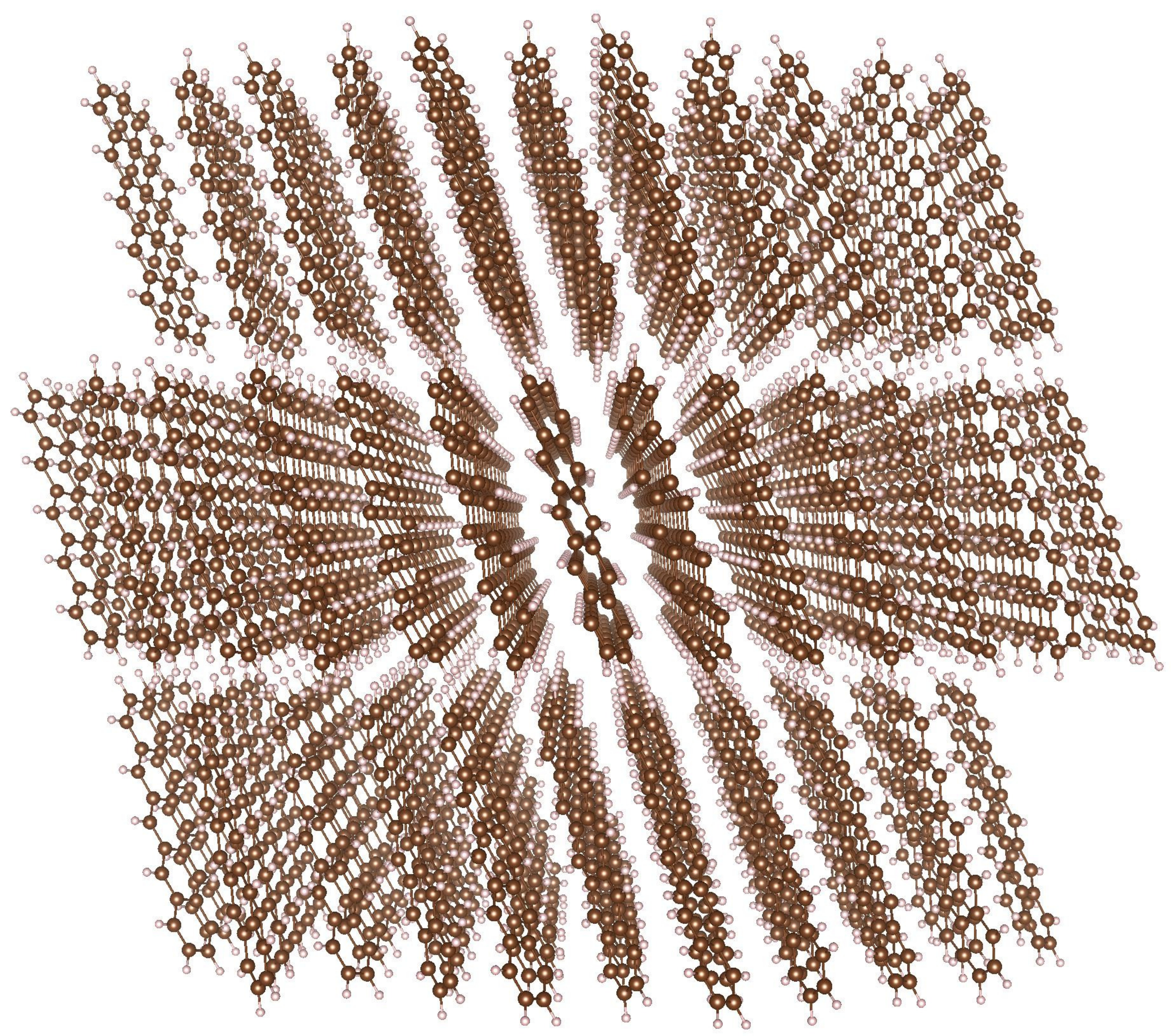}
  \label{fig:pentacene}}
  \hspace{10pt}
 \subfloat[][perovskite \ce{(PbI3CNH6)64}\newline(768 atoms)]{
  \includegraphics[width=0.15\textwidth]{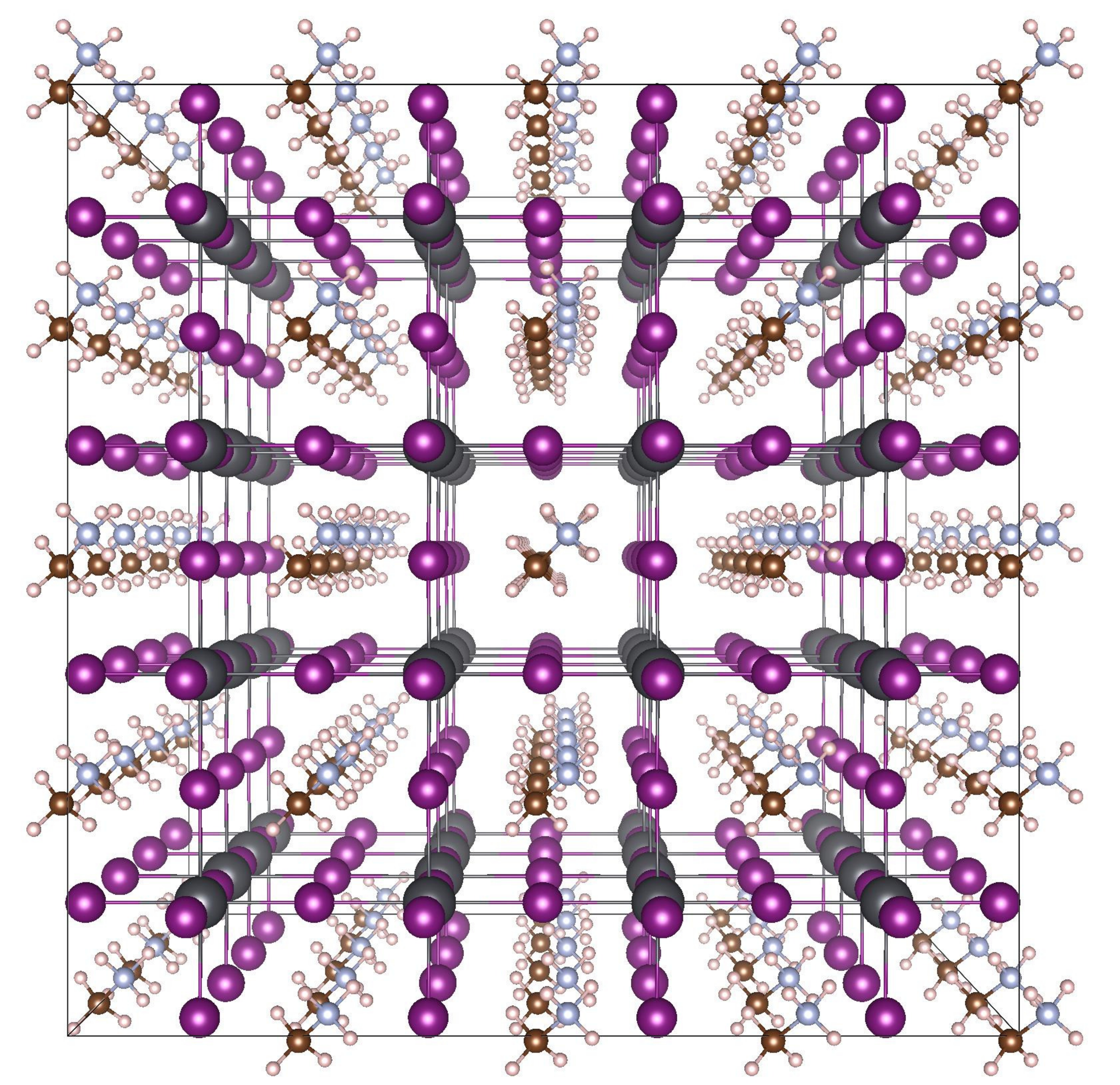}
  \label{fig:perovskite}}
  \hspace{10pt}
 \subfloat[][Si-wire\newline(706 atoms)]{
  \includegraphics[width=0.09\textwidth]{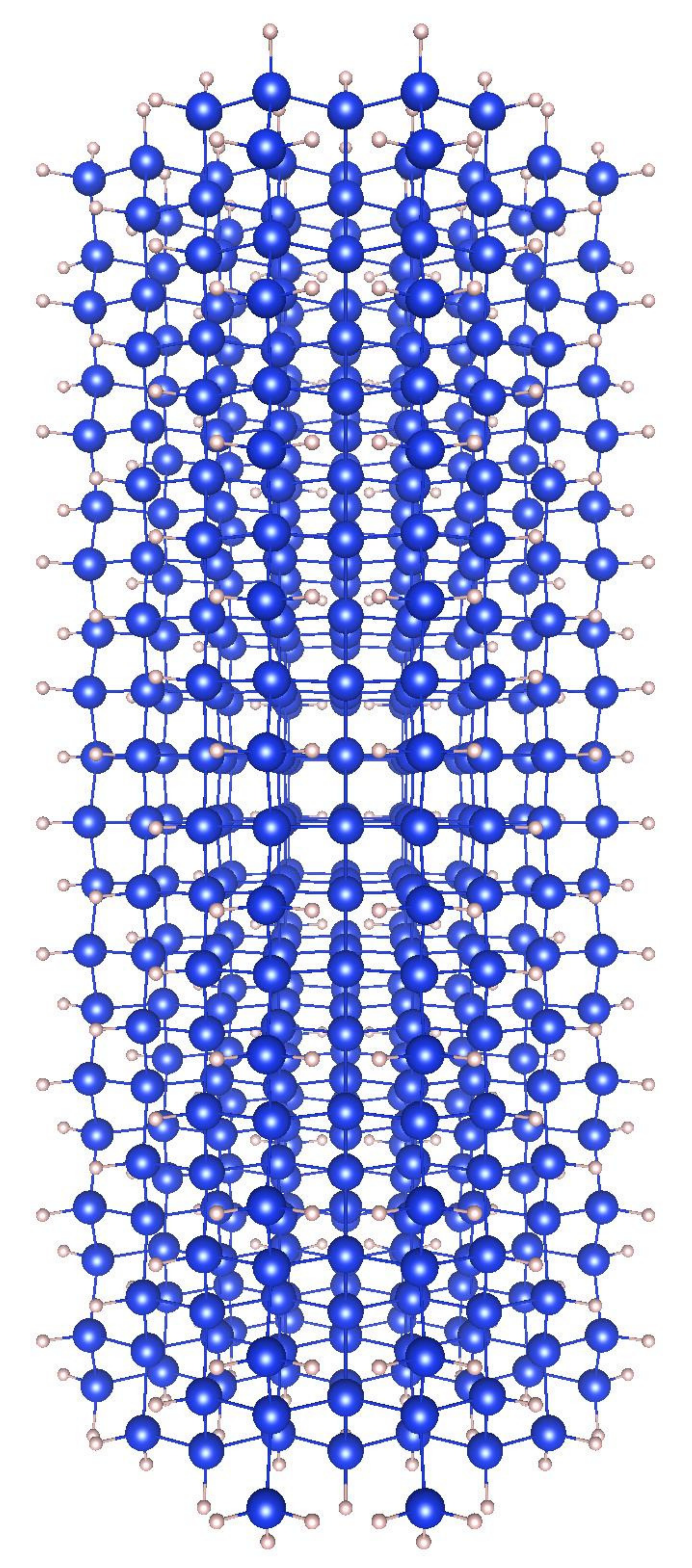}
  \label{fig:Si-wire}}
  \hspace{10pt}
 \subfloat[][water\newline(1800 atoms)]{
  \includegraphics[width=0.145\textwidth]{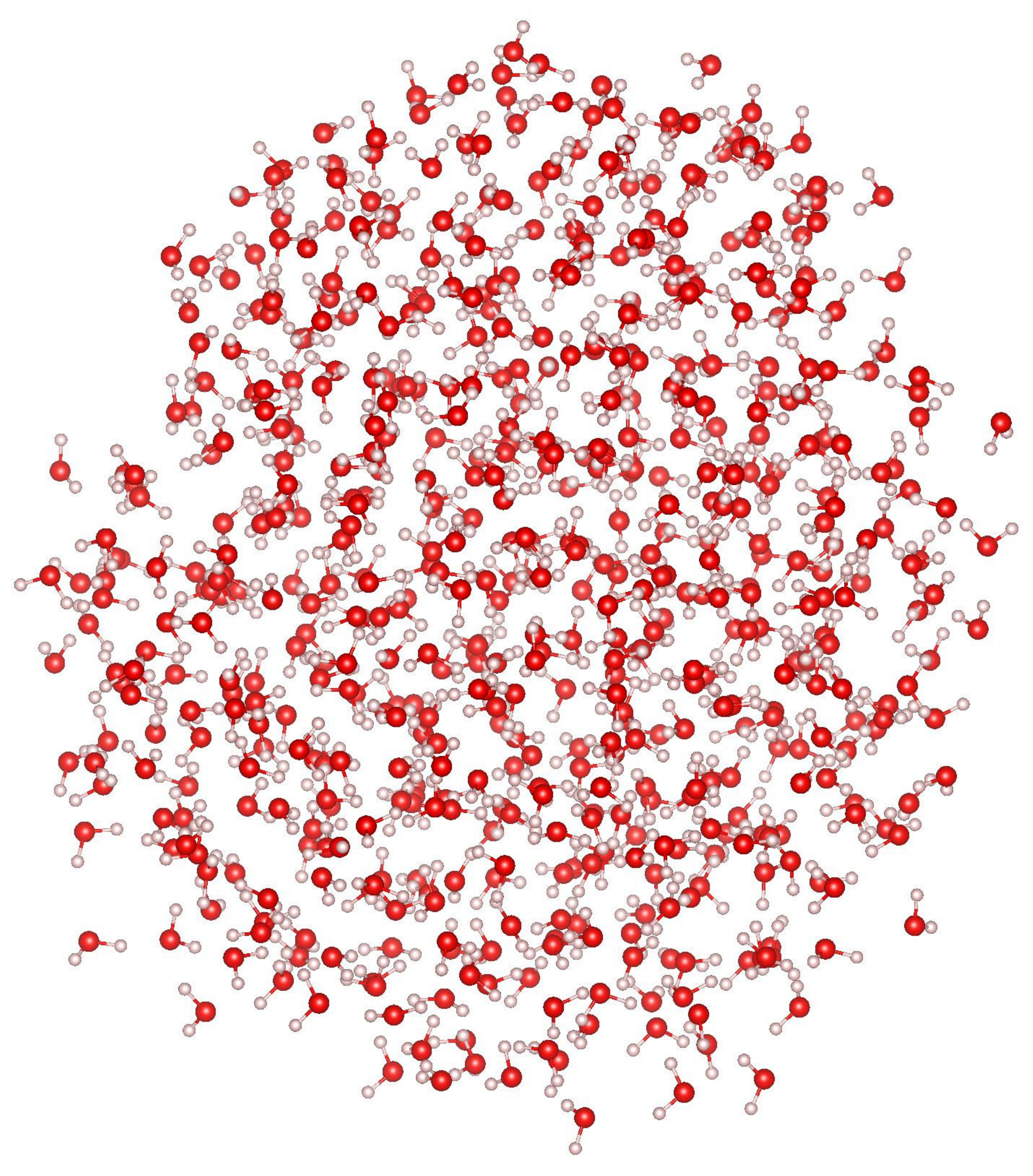}
  \label{fig:water}}
 \caption{The systems used for the analysis of the matrices produced by \textsc{BigDFT}; their data are shown in Tab.~\ref{tab:kappa_and_specwidth_BigDFT}.
 }
 \label{fig:BigDFT-systems}
\end{figure*}

\begin{table*}\small 
\begin{tabular}{l c r c rrrr c rrrrr}
 \toprule
    &&&& \multicolumn{4}{c}{$\mathbf{S}$} && \multicolumn{5}{c}{$\mathbf{H}$} \\
         \cmidrule{5-8} \cmidrule{10-14} \\
 system  && \#atoms && sparsity & $\epsilon_{min}$ & $\epsilon_{max}$ & $\kappa$ && sparsity & $\epsilon_{min}$ & $\epsilon_{max}$ & $\lambda$ & $\Delta_{HL}$ \\
         \cmidrule{1-1} \cmidrule{3-3} \cmidrule{5-8} \cmidrule{10-14} \\
 DNA            && 15613 && 99.57\% & 0.72   & 1.65   &  2.29  && 98.46\%  & -29.58  & 19.67 & 49.25 & 2.76 \\
 bulk pentacene &&  6876 && 98.96\% & 0.78   & 1.77   &  2.26  && 97.11\%  & -21.83  & 20.47 & 42.30 & 1.03 \\
 perovskite     &&   768 && 90.34\% & 0.70   & 1.50   &  2.15  && 76.47\% & -20.41  & 26.85 & 47.25 & 2.19 \\
 Si nanowire    &&   706 && 93.24\% & 0.72   & 1.54   &  2.16  && 81.61\% & -16.03  & 25.50 & 41.54 & 2.29 \\
 water          &&  1800 && 96.71\% & 0.83   & 1.30   &  1.57  && 90.06\%  & -26.55  & 11.71 & 38.26 & 9.95 \\
 \bottomrule
 \end{tabular}
 \caption{Sparsity, smallest and largest eigenvalue $\epsilon_{min}$ and $\epsilon_{max}$, and condition number $\kappa$ or spectral width $\lambda$, respectively, for the overlap and Hamiltonian matrix for typical runs with \textsc{BigDFT}. The eigenvalues shown for the overlap matrix are those of the standard eigenvalue problem $\mathbf{S}\mathbf{c}_i=\epsilon_i\mathbf{c}_i$, whereas in the case of the Hamiltonian matrix we report those of the generalized eigenvalue problem $\mathbf{H}\mathbf{c}_i=\epsilon_i\mathbf{S}\mathbf{c}_i$. For the Hamiltonian matrix we additionally show the HOMO-LUMO gap $\Delta_{HL}$. For the latter matrix all values are given in eV.}
 \label{tab:kappa_and_specwidth_BigDFT}
\end{table*}

As discussed in the introduction, there is a variety of methods to solve the typical problems arising in electronic structure theory and DFT in particular.
Consequently most approaches exhibit their best performance in a particular regime, determined by the properties of the matrices which are at the input of the problem.
In this section we briefly want to discuss the motivation for the creation of \textsc{CheSS}, i.e.\ present the conditions under which it works best.

As will be explained in more detail in Sec.~\ref{sec: Algorithm}, \textsc{CheSS} is based on a polynomial approximation, and hence the performance is a function on the polynomial degree.
The latter depends, first, on the specific function that has to be approximated and, second, on the interval over which the function shall be represented by the polynomial.
The first criterion is, of course, rather general and depends on the specific application; however, for electronic structure calculations a characteristic function is the Fermi function, which has the property of becoming less smooth (and hence harder to approximate) for systems with small band gaps at low electronic temperature.
Given this we can already now conclude that \textsc{CheSS} works best for systems exhibiting a decent gap between the Highest Occupied Molecular Orbital (HOMO) and the Lowest Unoccupied Molecular Orbital (LUMO), or for calculations with finite electronic temperature --- more details will be given later in Sec.~\ref{sec: Scaling with system properties}.
For the second point, namely the interval of the polynomial approximation, the situation is simpler:
In general, the larger the interval is the more polynomials will be required, 
since --- as explained in more detail later --- the necessary rescaling of the overall domain to the unit interval $[-1,1]$ results in higher polynomial degrees for the Chebyshev decomposition.
In our case, the interval of the approximation corresponds to the eigenvalue spectrum of the matrices --- the smaller this spectrum the better \textsc{CheSS} will perform.
Again we will investigate this point in more detail in Sec.~\ref{sec: Scaling with system properties}.

For DFT calculations the typical matrices that will be processed are the overlap matrix $S_{\alpha\beta}=\braket{\phi_\alpha|\phi_\beta}$ and the Hamiltonian matrix $H_{\alpha\beta}=\braket{\phi_\alpha|\mathcal{H}|\phi_\beta}$, where $\{\phi_\alpha\}$ is the used basis set and $\mathcal{H}$ the Hamiltonian operator.
The spectral width of $\mathbf{S}$ --- which we will from now on measure with the condition number $\kappa=\frac{\epsilon_{max}}{\epsilon_{min}}$, defined as the ratio between the largest and smallest eigenvalue, since the matrix is always positive definite
--- depends solely on the basis set.
Obviously, small and (quasi-)orthogonal basis sets are better suited than large non-orthogonal ones.
With respect to $\mathbf{H}$,
the spectrum depends as well on the basis; for instance, large basis sets include more high energetic states than small ones.
However, there is also a dependence on the physical model that is used, for instance whether the low energetic cores states are absorbed into a pseudopotential or not.
Since $\mathbf{H}$ has in general both negative and positive eigenvalues, the condition number is not a good measurement any more, and we therefore rather consider the total spectral width $\lambda=\epsilon_{max}-\epsilon_{min}$.

To see whether optimal conditions for \textsc{CheSS} --- i.e.\ a small spectral width for both $\mathbf{H}$ and $\mathbf{S}$ --- can be satisfied in practice, we investigated the properties of these two matrices for some typical calculations with \textsc{BigDFT}.
This code~\cite{mohr-2014-daubechies,mohr-2015-accurate} uses a \emph{minimal} set of \emph{quasi-orthogonal and in-situ optimized} basis functions.
The first property allows to keep the condition number of the overlap matrix small, whereas the second property --- together with the fact that \textsc{BigDFT} uses pseudopotentials to handle the core electrons~\cite{hartwigsen-relativistic-1998,willand-norm-conserving-2013} --- allows to operate with a Hamiltonian exhibiting a small spectrum.

As an illustration, we show in Tab.~\ref{tab:kappa_and_specwidth_BigDFT} the detailed values of the eigenvalue spectrum for typical runs with \textsc{BigDFT}, using systems that were already used in other publications~\cite{mohr-2015-accurate,mohr-fragments1-2017,mohr-fragments2-2017} and that are visualized in Fig.~\ref{fig:BigDFT-systems}.
As mentioned, \textsc{CheSS} is mainly designed for systems exhibiting a finite HOMO-LUMO gap, and therefore we only chose examples belonging to this class of systems.
As can be seen, the condition number is of the order of 2, independently of the system.
The same also holds for the spectral width of the Hamiltonian matrix, which is of the order of \unit[40-50]{eV}.
These low values are a direct consequence of the particular features of the \textsc{BigDFT} setup mentioned above.
For other popular basis sets, the condition numbers are usually considerably higher, even in case the basis sets were specifically designed to exhibit low values for $\kappa$, as for instance in the case of atomic orbitals (about two orders of magnitude larger)~\cite{berghold-polarized-2002} or Gaussians (at least one order of magnitude larger)~\cite{vandevondele-gaussian-2007}.
Thus, the fact that such low values can be reached within a DFT code illustrates the need for an algorithm that can exploit this feature and thus lead to very efficient calculations,
and indeed \textsc{CheSS} is used with great success together with \textsc{BigDFT}.
Moreover a low condition number of the overlap matrix is also crucial in the context of linear scaling algorithms, since it can be shown~\cite{maslen-locality-1998} that a localized and well-conditioned overlap matrix leads to an inverse with similar decay properties and finally also to an equally localized density matrix.

\subsection{Algorithm}
\label{sec: Algorithm}
\subsubsection{Expansion in Chebyshev polynomials}
\label{sec: Expansion in Chebyshev polynomials}
The basic ansatz of the algorithm behind \textsc{CheSS} is to approximate the matrix function $f(\mathbf{M})$ as a polynomial in $\mathbf{M}$.
However, such a polynomial expansion can become unstable for large degrees, which is known as Runge's phenomenon.
A way to circumvent this is to use Chebyshev polynomials, which are known to minimize this issue. 
This way the polynomial approximation becomes
\begin{equation}
  f(\mathbf{M}) \approx p(\mathbf{M}) = \frac{c_0}{2}\mathbf{I} + \sum_{i=1}^{n_{pl}} c_i \mathbf{T}^i(\mathbf{M}) \;,
 \label{eq:Chebyshev-polynomial}
\end{equation}
where $\mathbf{I}$ is the identity matrix 
and the $\mathbf{T}^i(\mathbf{M})$ are the Chebyshev matrix polynomials of order $i$.
Since these polynomials are only defined in the interval $[-1,1]$, the matrix $\mathbf{M}$ has to be scaled and shifted such that its eigenvalue spectrum lies within this range.
If $\epsilon_{min}$ and $\epsilon_{max}$ are the smallest and largest eigenvalue of $\mathbf{M}$, the modified matrix $\tilde{\mathbf{M}}$ that enters Eq.~\eqref{eq:Chebyshev-polynomial} is given by
\begin{equation}
 \tilde{\mathbf{M}} = \sigma(\mathbf{M}-\tau\mathbf{I}) \;,
 \label{eq:matrix_scaledandshifted}
\end{equation}
with
\begin{equation}
 \sigma=\frac{2}{\epsilon_{max}-\epsilon_{min}} \;, \quad \tau=\frac{\epsilon_{min}+\epsilon_{max}}{2} \;.
\end{equation}
Obviously, the eigenvalue spectrum of $\mathbf{M}$ is not known beforehand. However it is relatively easy to determine an approximate lower and upper bound 
 --- denoted as $\tilde{\epsilon}_{min}$ and $\tilde{\epsilon}_{max}$ --- 
for the eigenvalue spectrum on the fly, as will be shown later in Sec.~\ref{sec: Determination of the eigenvalue bounds and polynomial degree}.

The determination of the Chebyshev matrix polynomials and the expansion coefficients is straightforward~\cite{press-2007-numerical}. The polynomials can be calculated from the recursion relations
\begin{equation}
 \begin{split}
   & \mathbf{T}^0(\tilde{\mathbf{M}}) = \mathbf{I} \;, \\
   & \mathbf{T}^1(\tilde{\mathbf{M}}) = \tilde{\mathbf{M}}  \;, \\
   & \mathbf{T}^{j+1}(\tilde{\mathbf{M}}) = 2\tilde{\mathbf{M}}\mathbf{T}^j(\tilde{\mathbf{M}}) - \mathbf{T}^{j-1}(\tilde{\mathbf{M}}) \;,
 \end{split}
 \label{eq:Chebyshev-matrix-recursion}
\end{equation}
and the expansion coefficients are given by
\begin{multline}
 c_j = \frac{2}{n_{pl}} \times \\ \sum_{k=0}^{n_{pl}-1}f\left[\frac{1}{\sigma}\cos\left(\frac{\pi(k+\frac{1}{2})}{n_{pl}}\right)+\tau\right]\cos\left(\frac{\pi j(k+\frac{1}{2}}{n_{pl}}\right) \;, \\
 \label{eq:Chebyshev_expansion_coefficients_generalized}
\end{multline}
where $f(x)$ is the function that shall be applied to the matrix $\mathbf{M}$.

From Eq.~\eqref{eq:Chebyshev-matrix-recursion} it follows that the individual columns of the Chebyshev matrix polynomials fulfill as well a recursion relation and can be calculated independently of each other, i.e.\ we only need to apply local matrix-vector multiplications.
Eventually this also implies that each column of the matrix $p(\mathbf{M})$ can be calculated independently, which makes the algorithm highly efficient for large parallel computing architectures.
This specific need for sparse matrix-vector multiplications is in contrast to the more abundant case of parallel sparse matrix-matrix multiplications, for which various methods and libraries exist~\cite{rubensson-locality-aware-2016,weber-semiempirical-2015,borstnik-sparse-2014,bock-an-optimized-2013}.
In fact, there also exists the possibility to calculate the Chebyshev matrix polynomials using such matrix-matrix multiplications.
In this case, it is possible to reduce the required multiplications from $n_{pl}-1$ to about $2\sqrt{n_{pl}}$~\cite{liang-improved-2003}.
On the other hand, apart from loosing the strict independence of the columns and thus complicating the parallelization, this method has the additional drawback that some of the multiplications have to be repeated if the expansion coefficients change.
This is in contrast to our approach, where the individual columns can easily be summed up with different coefficients without the need of redoing any multiplications;
this feature is in particular important for the calculation of the density kernel, as will be shown in more detail in Sec.~\ref{sec:fermi_operator_expansion}.

\subsubsection{Determination of the eigenvalue bounds and polynomial degree}
\label{sec: Determination of the eigenvalue bounds and polynomial degree}

In order to get the estimates $\tilde{\epsilon}_{min}$ and $\tilde{\epsilon}_{max}$ for the eigenvalue spectrum on the fly,
we can use the same approach as outlined in Sec.~\ref{sec: Expansion in Chebyshev polynomials}.
Analogous to Eq.~\eqref{eq:Chebyshev-polynomial} we construct a penalty matrix polynomial $\mathbf{W}^p$,
where the expansion coefficients $c_i^{p}$ are again given by Eq.~\eqref{eq:Chebyshev_expansion_coefficients_generalized}, but with the function $f(x)$ being this time an exponential:
\begin{equation}
  f^{p}(x) = e^{\alpha(x-\tilde{\epsilon}_{min})} - e^{-\alpha(x-\tilde{\epsilon}_{max})}.
  \label{eq:penalty_function}
\end{equation}
If $\Tr({\mathbf{W}^p})$ is below a given numerical threshold, then all eigenvalues of $\mathbf{M}$ lie within the interval $[\tilde{\epsilon}_{min},\tilde{\epsilon}_{max}]$.
Otherwise, the trace will strongly deviate from zero, with the sign indicating which bound has to be adjusted.
The larger the value of $\alpha$, the more accurate the eigenvalue bounds can be determined, but on the other hand a higher degree of the Chebyshev expansion will be required to well represent this step-like function.
Since the calculation of the penalty matrix $\mathbf{W}^p$ uses the same Chebyshev polynomials as the original expansion of Eq.~\eqref{eq:Chebyshev-polynomial} and only requires to calculate a new set of expansion coefficients --- which is computationally very cheap --- this check of the eigenvalue bounds comes at virtually no extra cost and can therefore easily be done on the fly.
Nevertheless a good initial approximation of the eigenvalue spectrum is of course beneficial, as it avoids the recalculation of the polynomials in case the eigenvalue bounds have to be adjusted.
This is usually the case, since in a typical DFT setup \textsc{CheSS} is used within an iterative loop --- the so-called SCF cycle --- where the bounds only change little.
For the very first step, where no guess from a previous iteration is available, it is usually enough to start with typical default values for a given setup; 
in case that for some reason no such guess is available, it would still be possible to resort to other approaches, such as for instance a few steps of a Lanczos method~\cite{zhou-chebyshev-2014}, to get a reasonable starting value.

In order to determine automatically an optimal value for the polynomial degree $n_{pl}$,
we calculate a Chebyshev expansion $p_{m_{pl}}(x)$ for the one-dimensional function $f(x)$ --- which is computationally very cheap --- for various degrees $m_{pl}$, 
and then define the polynomial degree $n_{pl}$ as the minimal degree that guarantees that the polynomial approximation does not deviate from the function $f$ more than a given threshold $\lambda$:
 \begin{equation}
  n_{pl} = \min \big\{ m_{pl} \big| |p_{m_{pl}}(x)-f(x)|_{max} < \lambda \big\} \; .
 \end{equation}
Apart from the obvious dependence on the function $f$ that shall be represented, the minimal polynomial degree is also strongly related to the spectral width that must be covered.
In general $n_{pl}$ is smaller the narrower the spectral width is.
However, there can also be situations where it is advantageous to artificially spread the spectrum, as we briefly illustrate for the important case of the inverse.
Assuming that we have a matrix with a suitable condition number, but eigenvalues close to zero --- for instance a spectrum ranging from $10^{-2}$ to $1$ ---, a quite high degree will be required to accurately reproduce the divergence of the function $x^{-1}$ close to 0.
This problem can be alleviated by rescaling the matrix, which results in a larger spectral width, but yields a function that is easier to represent.
\textsc{CheSS} automatically detects such situations and rescales the matrix such that its spectrum lies further away from the problematic regions.

\subsection{Available operations}
\label{sec: Available operations}
\subsubsection{Fermi Operator Expansion}
\label{sec:fermi_operator_expansion}
A quantum mechanical system can be completely characterized by the density matrix operator $\hat{F}$, as the measure of any observable $\hat{O}$ is given by $\Tr(\hat{F}\hat{O})$.
Within DFT, it can be defined in terms of the eigenfunctions $\psi_i(\mathbf{r})$ --- which are solution of the single particle Schr\"odinger equation $\mathcal{H}\psi_i(\mathbf{r})=\epsilon_i\psi_i(\mathbf{r})$ --- as 
\begin{equation}
 \hat{F}(\mathbf{r},\mathbf{r}') = \sum_i f_i \psi_i(\mathbf{r})\psi(\mathbf{r}') \; ,
 \label{eq:density_matrix}
\end{equation}
with $f_i \equiv f(\epsilon_i)$ being the occupation of state $i$, determined by the Fermi function
\begin{equation}
 f(\epsilon) = \frac{1}{1+e^{\beta(\epsilon-\mu)}} \;,
 \label{eq:fermi_function}
\end{equation}
where $\mu$ is the Fermi energy and $\beta$ the inverse electronic temperature.
After choosing a specific basis $\{\phi_\alpha(\mathbf{r})\}$ --- i.e.\ $\psi_i(\mathbf{r}) = \sum_\alpha c_{i\alpha}\phi_\alpha(\mathbf{r})$ ---,
the Schr\"odinger equation becomes $\mathbf{H}\mathbf{c}_i=\epsilon_i\mathbf{S}\mathbf{c}_i$, and Eq.~\eqref{eq:density_matrix} corresponds to the calculation of the density kernel matrix $\mathbf{K}$,
\begin{equation}
 K _{\alpha\beta}= \sum_i f_i c_{i\alpha}c_{i\beta} \;.
\end{equation}
The drawback of this straightforward approach based on a diagonalization is that it scales cubically with respect to the size of the matrices and must therefore be avoided for the calculation of large systems.

The Fermi Operator Expansion (FOE)~\cite{goedecker-1994-efficient,goedecker-1995-tight-binding} that is implemented in \textsc{CheSS}
aims at calculating the density kernel directly from the Hamiltonian matrix (i.e.\ without a diagonalization), making the ansatz $\mathbf{K}=f(\mathbf{H})$.
In practice we can replace the Fermi function of Eq.~\eqref{eq:fermi_function} by any other function as long as it fulfills the essential feature of assigning an occupation of 1 to the occupied states and 0 to the empty states.
In our implementation we chose the complementary error function, since it decays rapidly from 1 to 0 around the Fermi energy:
\begin{equation}
 f(\epsilon) = \frac{1}{2} \left[ 1-\erf \big( \beta (\epsilon-\mu) \big) \right] \; .
 \label{eq:error_function}
\end{equation}
With these definitions the density kernel can then be calculated as outlined in Sec.~\ref{sec: Algorithm},
with the subtlety that the input matrix $\tilde{\mathbf{M}}$ has to be replaced by $\tilde{\mathbf{M}}'=\mathbf{S}^{-1/2} \tilde{\mathbf{M}} \mathbf{S}^{-1/2}$, and the output matrix has to be postprocessed as $\mathbf{S}^{-1/2}p(\tilde{\mathbf{M}}')\mathbf{S}^{-1/2}$.
The calculation of $\mathbf{S}^{-1/2}$ is as well done with \textsc{CheSS}, as will be explained in Sec.~\ref{sec: Matrix powers}.

Since the resulting density kernel must fulfill the condition $\Tr(\mathbf{K}\mathbf{S})=N$, where $N$ is the total number of electrons of the system, the parameter $\mu$ has to be adjusted until this condition is satisfied.
To do so we simply have to reevaluate Eq.~\eqref{eq:Chebyshev-polynomial} with a different set of coefficients, without the need of recalculating the Chebyshev polynomials, and this operation is consequently rather cheap.
The choice of $\beta$ is more delicate: It has to be chosen such that the error function decays from 1 to 0 within the range between the highest occupied state (i.e.\ the first state with an energy smaller than $\mu$) and the lowest unoccupied state (i.e.\ the first state with an energy larger than $\mu$).
Since this value is not known beforehand, we have to determine $\beta$ on the fly:
After calculating $\mathbf{K}$ with a first guess for $\beta$ we calculate a second kernel $\mathbf{K}'$ with a slightly larger decay length $\beta'>\beta$.
Then we compare the energies calculated with these two kernels: If the difference between $E=\Tr(\mathbf{K}\mathbf{H})$ and $E'=\Tr(\mathbf{K'}\mathbf{H})$ is below a given threshold, the decay length was sufficient; otherwise the density kernel has to be recalculated with a smaller value for $\beta$.
Usually this means that $n_{pl}$ must be increased, and thus a new set of polynomials must be calculated.
The second kernel $\mathbf{K}'$, on the other hand, can be evaluated cheaply as only the set of expansion coefficients in Eq.~\eqref{eq:Chebyshev-polynomial} changes.

\subsubsection{Matrix powers}
\label{sec: Matrix powers}
The calculation of matrix powers, i.e.\ $\mathbf{M}^a$,
can be done exactly along the same lines as the calculation of the density kernel described in Sec.~\ref{sec:fermi_operator_expansion},
with the only difference that the function that enters the calculation of the Chebyshev expansion coefficients of Eq.~\eqref{eq:Chebyshev_expansion_coefficients_generalized} is now given by $f(x)=x^a$.
Our approach allows to calculate any --- also non-integer --- power $a$, as long as the function $f$ can be well represented by Chebyshev polynomials throughout the entire eigenvalue spectrum.
Depending on the power $a$ this might lead to some restrictions;
for the inverse for instance this means that the matrix should be positive definite, since otherwise the divergence at $x=0$ will lead to problems.
For typical applications within electronic structure codes, the matrix $\mathbf{M}$ for which powers have to be calculated is the overlap matrix, and hence the above requirement is always fulfilled.

\subsubsection{Extraction of selected eigenvalues}
The FOE formalism described in Sec.~\ref{sec:fermi_operator_expansion} can also be used to extract selected eigenvalues from a matrix.
To this end one has to recall that for a system with total occupation $\Tr(\mathbf{KS})=N$ the Fermi energy will be chosen such that the error function of Eq.~\eqref{eq:error_function} decays from 1 to 0 between the $N$th and $(N+1)$th eigenvalue.
Therefore, in order to extract the $N$th eigenvalue, we simply perform an FOE calculation using a total occupation of $\Tr(\mathbf{KS})=N-\frac{1}{2}$.
This will lead to an error function that decays in such a way that it only half populates the $N$th eigenstate, which is the case if the Fermi energy coincides with the $N$th eigenvalue.
The accuracy of the calculated eigenvalue is related to the value of $\beta$; we will discuss this in more detail in Sec.~\ref{sec: Accuracy}.

Even though this approach to extract eigenvalues is originally related to the calculation of the density kernel, it is 
generally applicable to any matrix as long as
the error function of Eq.~\ref{eq:error_function} is well representable by Chebyshev polynomials over the entire eigenvalue spectrum.
Moreover, the Chebyshev polynomials --- whose calculation is the most expensive part of this approach --- can be reused for each eigenvalue, since only the value of $\mu$ has to be varied.

Nevertheless we should see this eigenvalue calculation more as a ``side product'' of the usage of \textsc{CheSS} rather than an individual feature.
In other terms, once the Chebyshev polynomials are determined --- for instance to calculate the density kernel --- a rough estimate of some eigenvalues (e.g.\ the HOMO-LUMO gap) can be calculated using the very same polynomials with hardly any additional cost.
If, on the other hand, the accurate (and not just approximate) calculation of the eigenvalues and associated eigenvectors is a central point of an algorithm, then one should most likely resort to more appropriate interior eigenvalue solvers, such as for instance the shift-and-invert Lanczos method~\cite{ericsson-the-spectral-1980}, the Sakurai-Sugiura method~\cite{sakurai-a-projection-2003} or the FEAST algorithm~\cite{polizzi-density-matrix-based-2009}.
An example for such an approach is shown in Ref.~\citenum{nakata-efficient-2017}, where the Sakurai-Sugiura method~\cite{sakurai-a-projection-2003} is used to compute hundreds of interior eigenstate for a Hamiltonian matrix stemming from a large-scale DFT calculation with the \textsc{Conquest} code~\cite{bowler-practical-2000,bowler-recent-2006,bowler-an-overview-2010}.


\subsection{Sparsity and Truncation}
\label{sec: Sparsity and truncation}
\paragraph{Truncation to a fixed sparsity pattern}
\textsc{CheSS} is designed for large sparse matrices, meaning that most of the matrix entries are zero and consequently not stored.
However, applying an operation to a sparse matrix --- in our case the matrix vector multiplications to build up the Chebyshev polynomials 
--- does in general not preserve this sparsity, and repeated application of this operation can eventually lead to a dense matrix.
To avoid this, the result after applying the operation is again mapped onto a sparsity pattern, i.e.\ certain entries are forced to be zero.
There are various ways to enforce such a sparsity, for instance simple methods such as setting all elements below a given threshold value to zero or using information about the distance of the basis functions corresponding to a given matrix entry, as well as more sophisticated approaches that allow to control the error introduced by the truncation~\cite{rubensson-systematic-2005}.
In our case, we work with a fixed sparsity pattern for the sake of simplicity.
In order to keep the error introduced by the truncation small, the sparsity pattern of the matrix after the applied operation has to be slightly larger than the original one, meaning that there must be some ``buffers'' into which the matrix can extend during the applied operation.
The size of these buffers is related to the properties of the matrix and the specific operation, and hence the correct setup depends on the particular application.

\begin{figure}
 \centering
 \subfloat[][original matrix $\mathbf{M}$  \label{fig:sparsity-pattern-original_full}]{\includegraphics[width=0.229\textwidth]{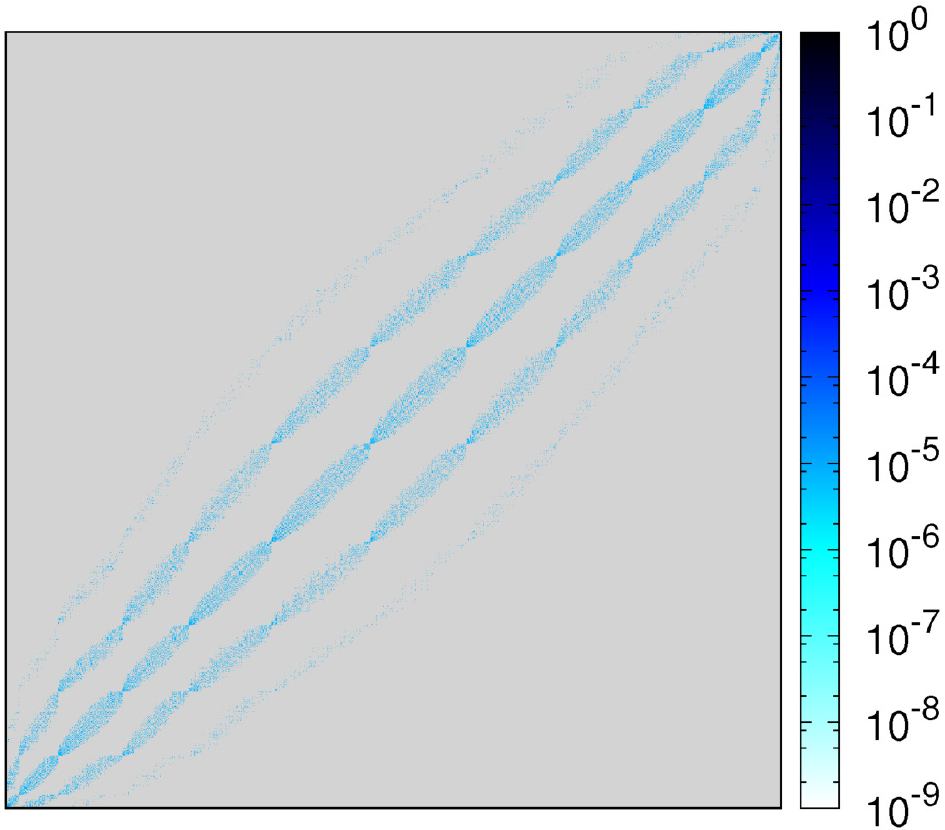}} \hspace{6pt}
 \subfloat[][exact calculation of $\mathbf{M}^{-1}$ without sparsity constraints \label{fig:sparsity-pattern-inverse_full}]{\includegraphics[width=0.229\textwidth]{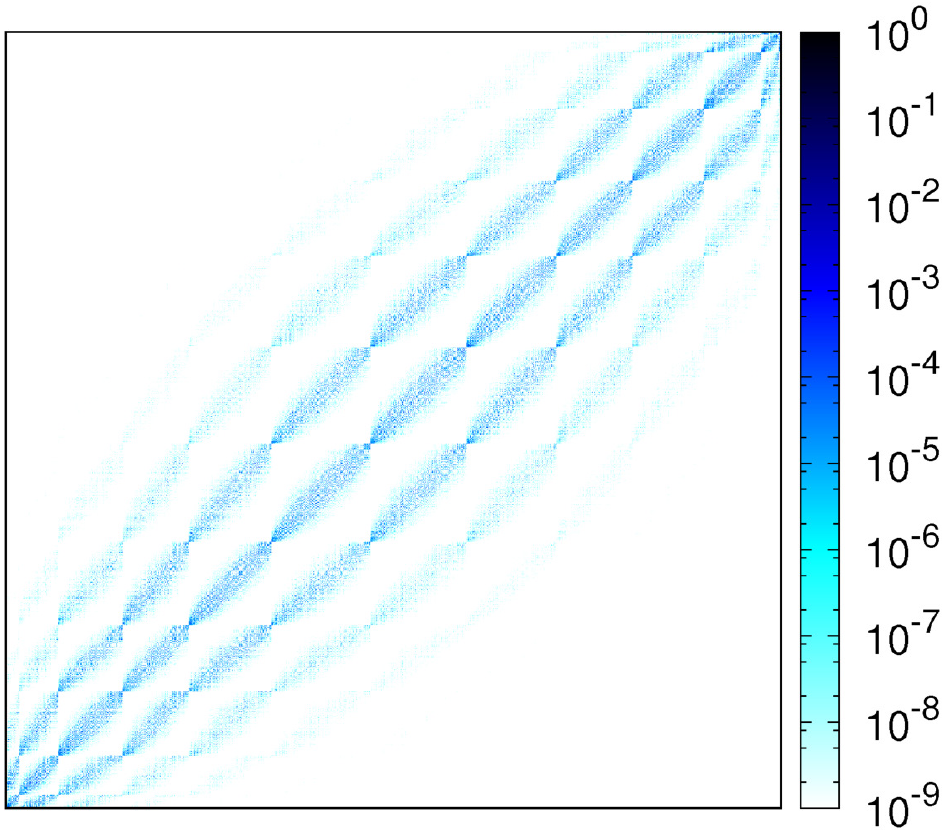}} \\
 \subfloat[][sparse calculation of $\mathbf{M}^{-1}$ using \textsc{CheSS} within the sparsity pattern \label{fig:sparsity-pattern-inverse_sparse}]{\includegraphics[width=0.229\textwidth]{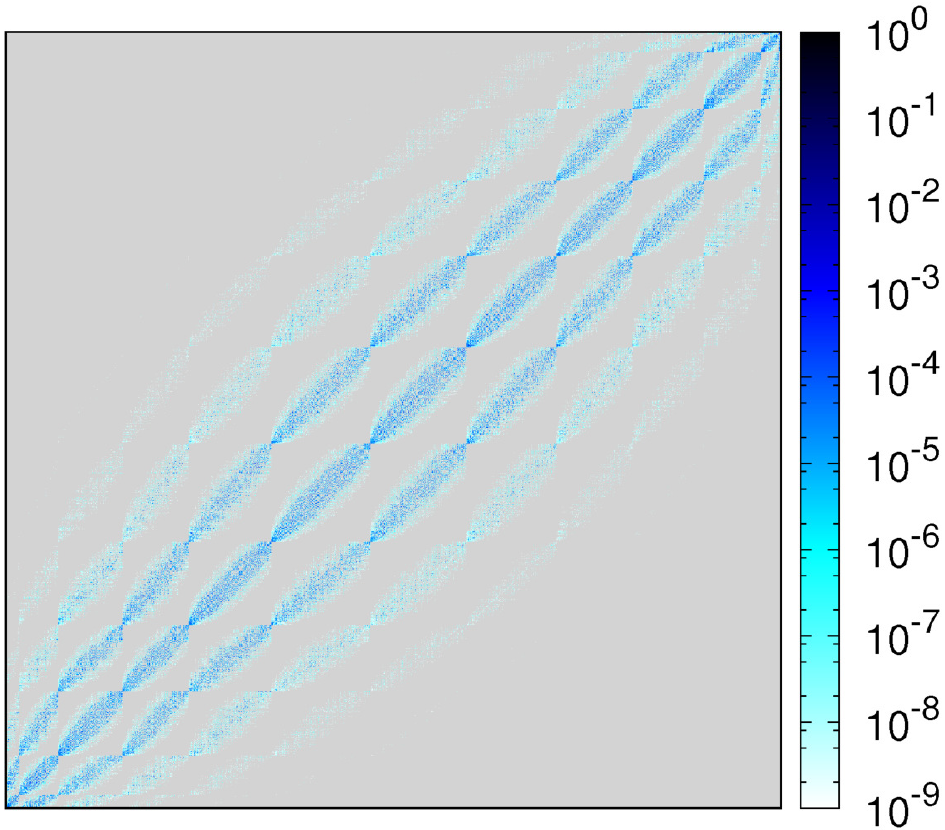}} \hspace{6pt}
 \subfloat[][difference between Fig.~\ref{fig:sparsity-pattern-inverse_full} and Fig.~\ref{fig:sparsity-pattern-inverse_sparse} \label{fig:sparsity-pattern-inverse_difference}]{\includegraphics[width=0.229\textwidth]{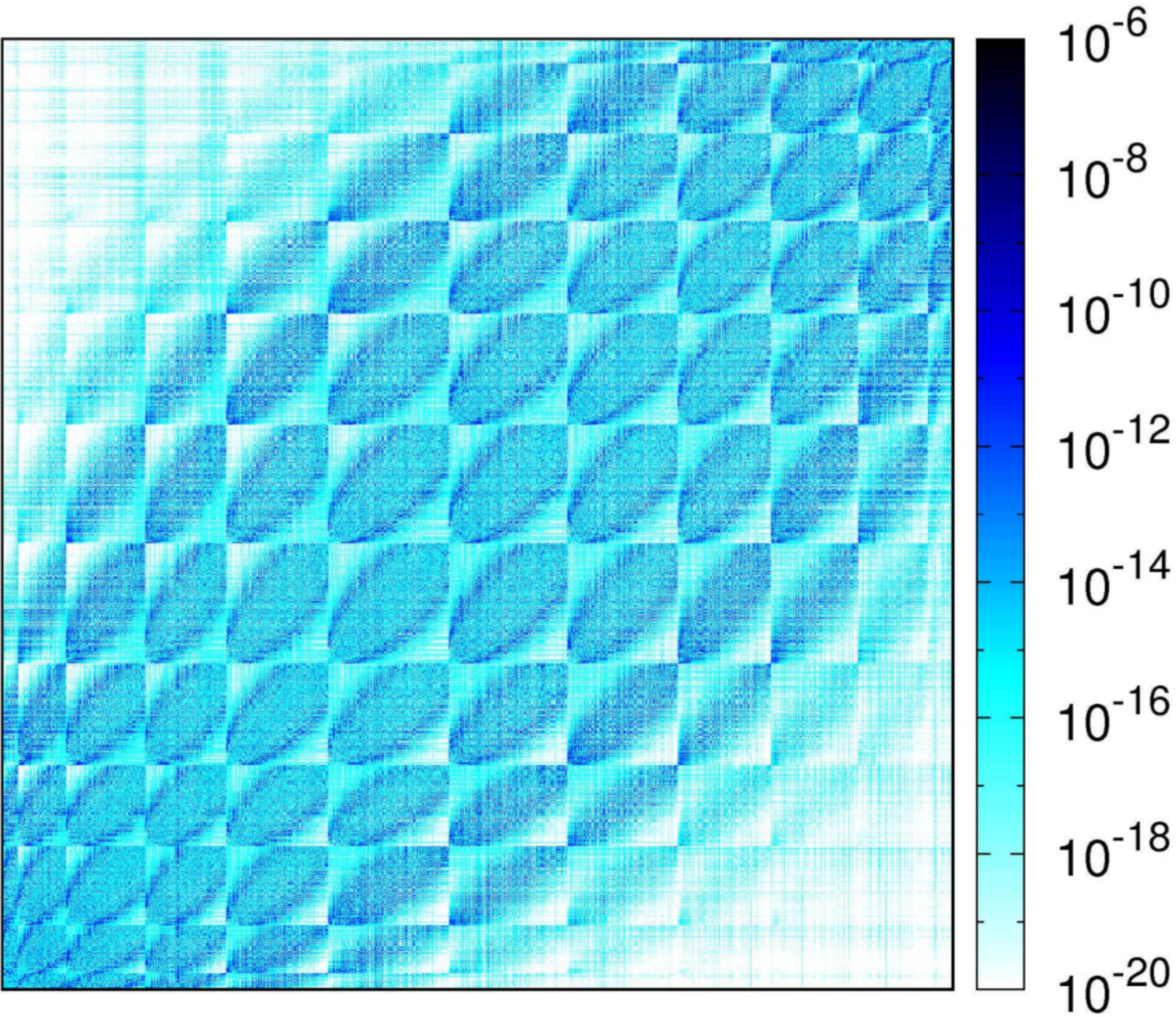}}
 \caption{Heat map of the matrix elements for a $6000 \times 6000$ sparse matrix $\mathbf{M}$ and its inverse $\mathbf{M}^{-1}$. 
 The sparsity pattern of the matrices originate from a calculation of a small water droplet with BigDFT, and contain $738,316$ non-zero elements ($97.95$\% sparsity) for $\mathbf{M}$ and --- due to the buffer regions --- $4,157,558$ non-zero elements ($88.45\%$ sparsity) for $\mathbf{M}^{-1}$.
 We filled the matrix with random numbers, in such a way that the matrix is symmetric and positive definite, and with eigenvalues in the range from $0.1$ to $1.1$, corresponding to a condition number of $11$. 
 Values that are zero due to the sparsity pattern are marked in gray. 
 For the sake of a better visualization the coloring scheme ends at $10^{-9}$ and $10^{-20}$, respectively, even though there are still much smaller values.
  }
 \label{fig:sparsity-pattern}
\end{figure}

As a small illustration we show in Fig.~\ref{fig:sparsity-pattern} the behavior for the calculation of the inverse.
In Fig.~\ref{fig:sparsity-pattern-original_full} we show the values of the original matrix, with the entries that are strictly zero due to the sparsity pattern marked in gray. In Fig.~\ref{fig:sparsity-pattern-inverse_full} we show the inverse, calculated exactly and without any sparsity constraint; as can be seen, this matrix is less sparse than the original one, but nevertheless far from being dense.
Consequently, it is reasonable to again map its values onto a sparsity pattern.
However, due to the larger extent, this sparsity pattern has to contain the aforementioned buffer regions.
This is illustrated in Fig.~\ref{fig:sparsity-pattern-inverse_sparse}, where we show the inverse calculated by \textsc{CheSS} within the predefined enlarged sparsity pattern.
Indeed we see that this pattern has been chosen reasonably and is able to absorb the spread of the inverse compared to the original matrix.
This becomes even better visible in Fig.~\ref{fig:sparsity-pattern-inverse_difference}, where we show --- using a different scale for the colors --- the absolute difference between the exact solution and the one calculated by \textsc{CheSS}.
We show this difference for the entire matrix, i.e.\ both inside and outside of the sparsity pattern;
as can be seen the difference is very small throughout the entire matrix --- the maximal difference is only $1.2 \times 10^{-7}$ ---,
showing firstly that the buffer region has been chosen reasonably and secondly that the inverse within the sparsity pattern has been accurately calculated by \textsc{CheSS}.

\paragraph{Effects on the error definitions}
The fact that the resulting matrix is mapped back onto a sparsity pattern has also an impact on the definition of the ``exact solution''. 
There are two ways to define the exact sparse solution of an arbitrary operation $f(\mathbf{M})$:
\begin{enumerate}
 \item By defining the exact solution as the one that we obtain by calculating $f(\mathbf{M})$ without any constraints and then cropping the result to the desired sparsity pattern. The drawback of this definition is that it in general violates the essential identity $f^{-1}(f(\mathbf{M}))=\mathbf{M}$.
 \item By calculating the solution directly within the sparsity pattern, symbolized as $\hat{f}(\mathbf{M})$, and then defining the exact solution as the one which fulfills $\hat{f}^{-1}(\hat{f}(\mathbf{M}))=\mathbf{M}$.
\end{enumerate}
\textsc{CheSS}
calculates the solution by construction within the predefined sparsity pattern, and the exact solution is therefore defined in the second way.
More quantitatively, the error according to this second definition is given by
\begin{multline}
 w_{\hat{f}^{-1}(\hat{f})} = \frac{1}{|\hat{f}(\mathbf{M})|} \times \\ \sqrt{ \sum_{(\alpha\beta) \in \hat{f}(\mathbf{M})} \left( \hat{f}^{-1}(\hat{f}(\mathbf{M}))_{\alpha\beta}-M_{\alpha\beta} \right)^2 } \; ,
 \label{eq:error_definition_inverse}
\end{multline}
where $\sum_{(\alpha\beta) \in \hat{f}(\mathbf{M})}$ indicates a summation over all elements within the predefined sparsity pattern of $\hat{f}(\mathbf{M})$ and $|\hat{f}(\mathbf{M})|$ denotes the number of elements within this pattern.
Nevertheless we will also report errors 
according to the first definition, in order to asses the effect of the sparsity pattern and the resulting truncation, and define this error as 
\begin{multline}
 w_{\hat{f}_{sparse}} = \frac{1}{|\hat{f}(\mathbf{M})|} \times \\ \sqrt{ \sum_{(\alpha\beta) \in \hat{f}(\mathbf{M})} \left( \hat{f}(\mathbf{M})_{\alpha\beta}-f(\mathbf{M})_{\alpha\beta} \right)^2 } \;.
 \label{eq:error_definition_sparse}
\end{multline}

\paragraph{Fixed versus variable sparsity}
Related to the topic of sparsity and truncation, we finally also want to point out an important advantage of our approach to calculate the density kernel compared to another popular class of methods, namely density matrix purification~\cite{mcweeny-the-1956,mcweeny-some-1960,palser-canonical-1998,holas-transforms-2001,niklasson-expansion-2002,niklasson-trace-2003,mazziotti-towards-2003,xiang-spin-unrestricted-2005,rubensson-systematic-2005}.
Both methods, i.e.\ FOE and purification, are based on a series of repeated matrix multiplications;
however, whereas the FOE method directly expands the density matrix as a polynomial of the Hamiltonian, purification recursively applies low order polynomials.
As a result of this recursive procedure, these methods require fewer matrix multiplications than the FOE method.

\begin{figure}
 \includegraphics[width=1.0\columnwidth]{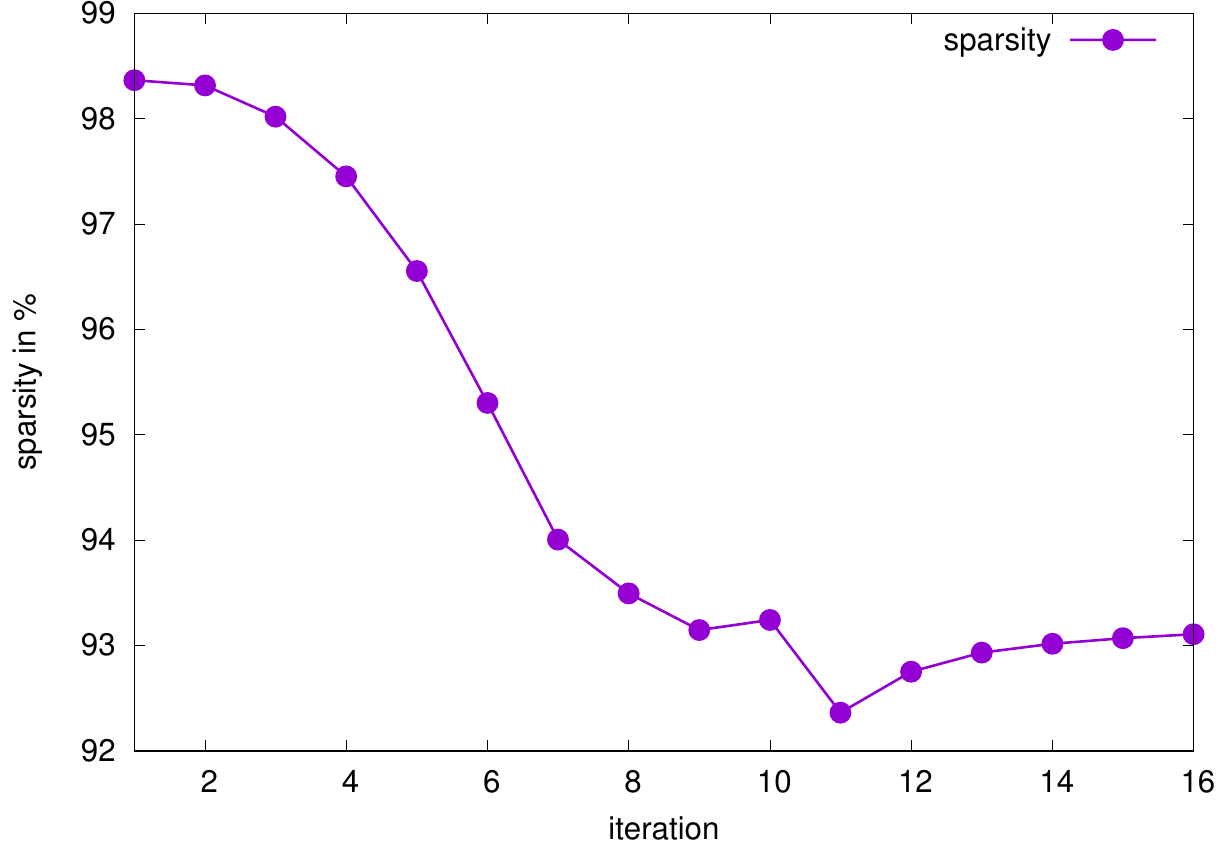}
 \caption{
  Fill-in of the intermediate matrices, represented by its sparsity, in the course of a purification calculation using the fourth-order trace resetting method (TRS4)~\cite{niklasson-expansion-2002}.
  As a test system we used a DNA fragment in solution (17947 atoms), giving rise to a matrix of size $36460 \times 36460$.
  }
 \label{fig:purification_sparsity}
\end{figure}

However, the cost of the multiplications performed in recursive procedures and the FOE method are not equal.
This is because of matrix fill-in.
In Fig.~\ref{fig:purification_sparsity}, we show the fill-in over the course of a representative purification calculation.
This calculation used the fourth-order trace resetting method (TRS4) of Niklasson~\cite{niklasson-expansion-2002}, and was performed on a system representing a DNA fragment in solution.
The sparsity is here defined in a variable way, namely by the magnitude of the matrix entries, i.e.\ all elements below a given threshold are set to zero.
Indeed we see that the sparsity decreases considerably from more than 98\% to about 93\%.
This fill-in is even greater for poor starting guesses, systems with smaller HOMO-LUMO gaps, or calculations requiring greater accuracy.

During a density matrix purification calculation, the performance is determined by the cost of multiplying two matrices that have as many nonzero elements as the final density matrix.
By contrast, \textsc{CheSS} always multiplies the intermediate density matrix by the sparser Hamiltonian matrix.
This allows to determine beforehand the (constant) cost of the matrix multiplications, whereas for the purification schemes it may explode unexpectedly.
Finally, the fact that in FOE we always apply the same matrix also permits a much easier parallelization --- each column of the final matrix can be calculated independently --- compared to the purification approach.
This, combined with the aforementioned strict sparsity, allows \textsc{CheSS} to perform accurate calculations with great efficiency, as will be demonstrated in section \ref{sec: Performance}.

\subsection{Storage format of the sparse matrices}
\label{sec: Storage format of the sparse matrices}
\begin{figure}
 \includegraphics[width=1.0\columnwidth]{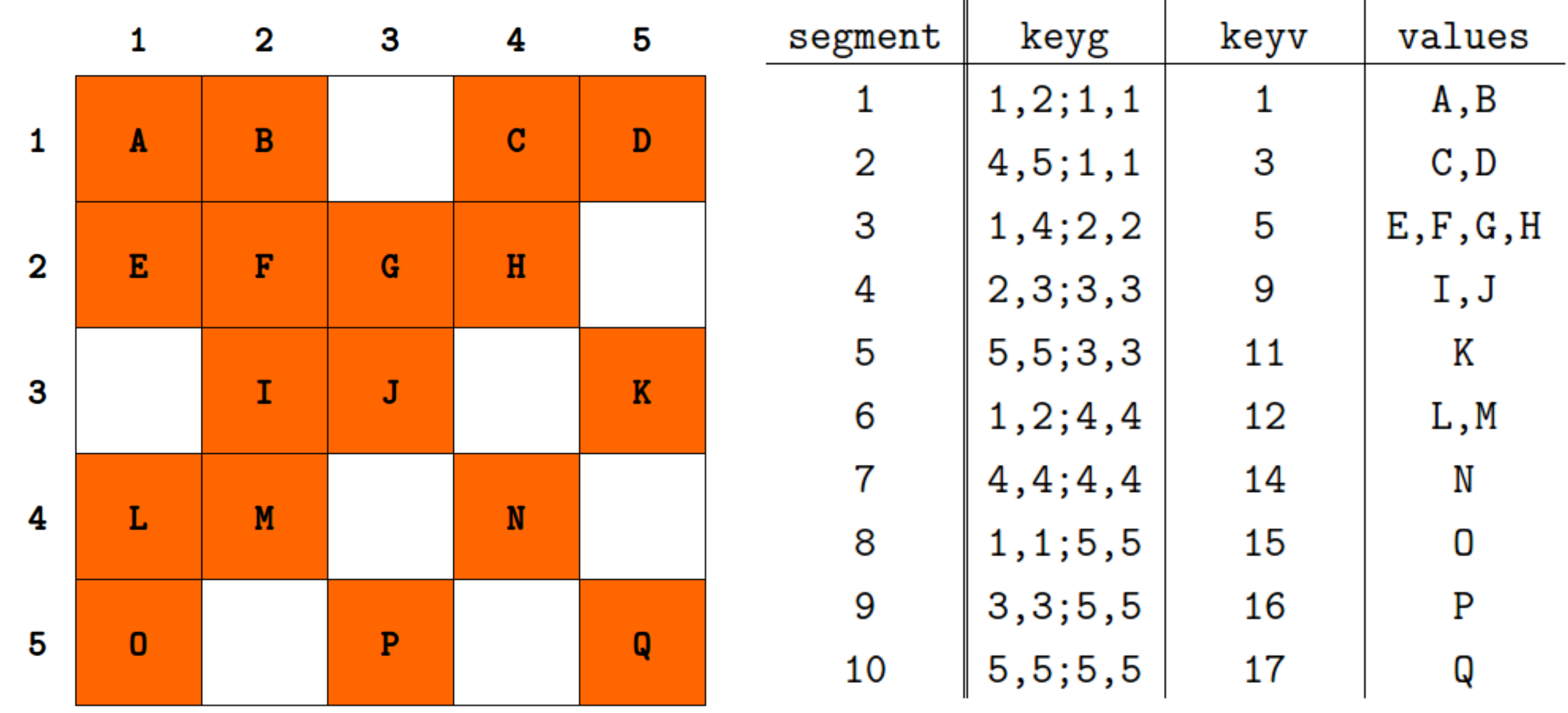}
 \caption{Schematic illustration of the descriptors \texttt{keyg} and \texttt{keyv} used by the SSF format to store a sparse matrix. Consecutive non-zero entries are grouped together in segments, which are in this toy example however restricted to a single row each. The values of each segment are simply stored in a consecutive one-dimensional array. The example shows the format for a row storage implementation, but the same concept is also applicable to a column storage setup.
 }
 \label{fig:toy_system}
\end{figure}

\textsc{CheSS} is designed to work with large sparse matrices and consequently only stores the non-zero entries within a single one-dimensional array.
To describe the corresponding sparsity pattern, it uses a special format that we denote as Segment Storage Format (SSF).
The basic idea of the SSF format is to group together consecutive nonzero entries as segments.
Assuming that we have $nseg$ such segments, the SSF format then requires two descriptor arrays, denoted by \texttt{keyg} and \texttt{keyv}, of dimension $(2,2,nseg)$ and $(nseg)$, respectively.
\texttt{keyg} indicates the start and end of each segment in ``dense coordinates'', i.e.\ each entry has the form $(c_s,c_e;r_s,r_e)$, with $c_s$ and $c_e$ denoting the starting and ending column, respectively, and $r_s$ and $r_e$ denoting the starting and ending row, respectively.
\texttt{keyv} indicates at which entry within the array of non-zero entries a given segment starts; this array actually contains redundant information that can be reconstructed at any time from \texttt{keyg} and is only used to accelerate the handling of the sparse matrices.
An illustration of this storage format for a simple $5\times5$ matrix is shown in Fig.~\ref{fig:toy_system}.

The advantage of this format is that it allows to describe the sparsity pattern in a very compact form.
For a matrix containing $nseg$ segments, it requires $5\times nseg$ descriptor elements, or actually only $4\times nseg$ if the redundant \texttt{keyv} descriptors are omitted.
The CCS/CRS format, which is a standard format to store sparse matrices, requires for the same description  $ncol+nnz$ elements, with $ncol$ being the number of columns/rows and $nnz$ being the number of non-zero entries. 
Assuming that $nnz \gg ncol$ and thus neglecting the contribution coming from the $ncol$ elements, we consequently see that out format is more compact as soon as the average segment size is larger than 4, which is likely to be the case for many applications.
This does not only reduce the memory footprint during a calculation, but also speeds up I/O operations and cuts down the required disk space for storage.

Even though \textsc{CheSS} is designed for large scale applications, the focus onto electronic structure methods --- which are computationally very expensive --- nevertheless limits the matrix sizes that are typically handled by the library.
Consequently it is in most cases not necessary to resort to complicated parallel distribution schemes for the sparse matrices.
Nevertheless we have implemented such a distribution scheme to make \textsc{CheSS} also usable in extreme situations. For technical details we refer to the appendix of Ref.~\citenum{mohr-2015-accurate}.

\section{Performance}
\label{sec: Performance}
In the following we will present various benchmarks in order to evaluate the accuracy and performance of \textsc{CheSS}.
The sparsity patterns of all matrices used for these tests are coming from calculations of small water droplets with the \textsc{BigDFT}~\cite{mohr-2014-daubechies,mohr-2015-accurate} code.
The buffer regions mentioned in Sec.~\ref{sec: Sparsity and truncation} are based on simple geometrical criteria;
nevertheless this does not decrease the validity of the following tests,
as the sparsity pattern is always something that depends on the specific application and is therefore determined by the code interfacing with \textsc{CheSS}.
Moreover, we took --- unless otherwise stated --- from the \textsc{BigDFT} runs only the sparsity pattern, but not the content of the matrices;
rather they were filled with random numbers in order get the desired properties, like for instance the spectral width.

For the case of the matrix powers, we focus on the important case of the inverse;
however this is not a restriction, as other powers can be calculated exactly along the same lines.
For the extraction of selected eigenvalues we do not show many performance data, as most would be redundant with the data shown for the calculation of the density kernel.

\subsection{Accuracy}
\label{sec: Accuracy}

In this section we want to assess the accuracy of \textsc{CheSS}, for each of the available operations.
In all cases we took as example a matrix of dimension $6000 \times 6000$, with a degree of sparsity of $97.95\%$ for $\mathbf{S}$, $92.97\%$ for $\mathbf{H}$, and $88.45\%$ for $\mathbf{S}^{-1}$ and $\mathbf{K}$.

\paragraph{Inverse}

\begin{figure}
 \includegraphics[width=1.0\columnwidth]{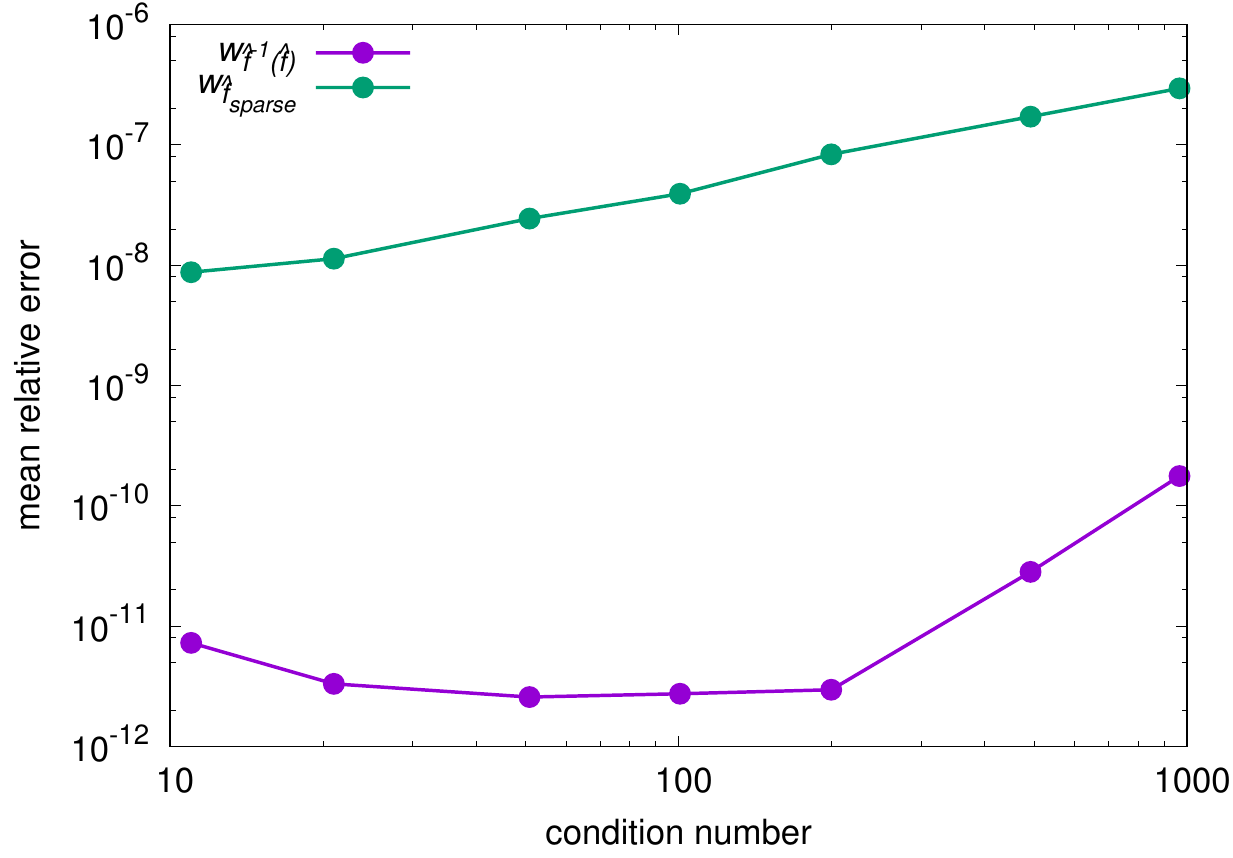}
 \caption{
 Mean error for the calculation of the inverse, according to Eq.~\eqref{eq:error_definition_inverse} (``$w_{\hat{f}^{-1}(\hat{f})}$'') and Eq.~\ref{eq:error_definition_sparse} (``$w_{\hat{f}_{sparse}}$''), however considering the relative error instead of the absolute error.
 In order to avoid divisions by zero only values larger than $10^{-12}$ were considered.
 }
 \label{fig:accuracy_ice}
\end{figure}

In Fig.~\ref{fig:accuracy_ice} we show the errors for the calculation of the inverse, according to Eqs.~\eqref{eq:error_definition_inverse} and \eqref{eq:error_definition_sparse}, as a function of the condition number.
However, in order to capture also differences between small numbers, we actually report the \emph{relative} error, i.e.\ we replace in Eq.~\eqref{eq:error_definition_inverse} $\left( \hat{f}^{-1}(\hat{f}(\mathbf{M}))_{\alpha\beta}-M_{\alpha\beta} \right)^2$ by $\left( \frac { \hat{f}^{-1}(\hat{f}(\mathbf{M}))_{\alpha\beta}-M_{\alpha\beta} } { M_{\alpha\beta} }\right)^2$ and in Eq.~\eqref{eq:error_definition_sparse} $\left( \hat{f}(\mathbf{M})_{\alpha\beta}-f(\mathbf{M})_{\alpha\beta} \right)^2$ by $\left( \frac { \hat{f}(\mathbf{M})_{\alpha\beta}-f(\mathbf{M})_{\alpha\beta} } { f(\mathbf{M})_{\alpha\beta} } \right)^2$.
The polynomial degree was determined automatically, as described in Sec.~\ref{sec: Determination of the eigenvalue bounds and polynomial degree}, with a value of $\alpha=-200$ for the penalty function of Eq.~\eqref{eq:penalty_function}; this leads to values $n_{pl}$ ranging from 60 to 4670, depending on the condition number.
As can be seen, the error according to Eq.~\eqref{eq:error_definition_inverse} is essentially zero,  confirming the accuracy of the Chebyshev fit.
The error according to Eq.~\eqref{eq:error_definition_sparse} is slightly larger, but nevertheless remains small for all values of the condition number, indicating that the buffer regions have been chosen sufficiently large.

\begin{figure}
 \includegraphics[width=1.0\columnwidth]{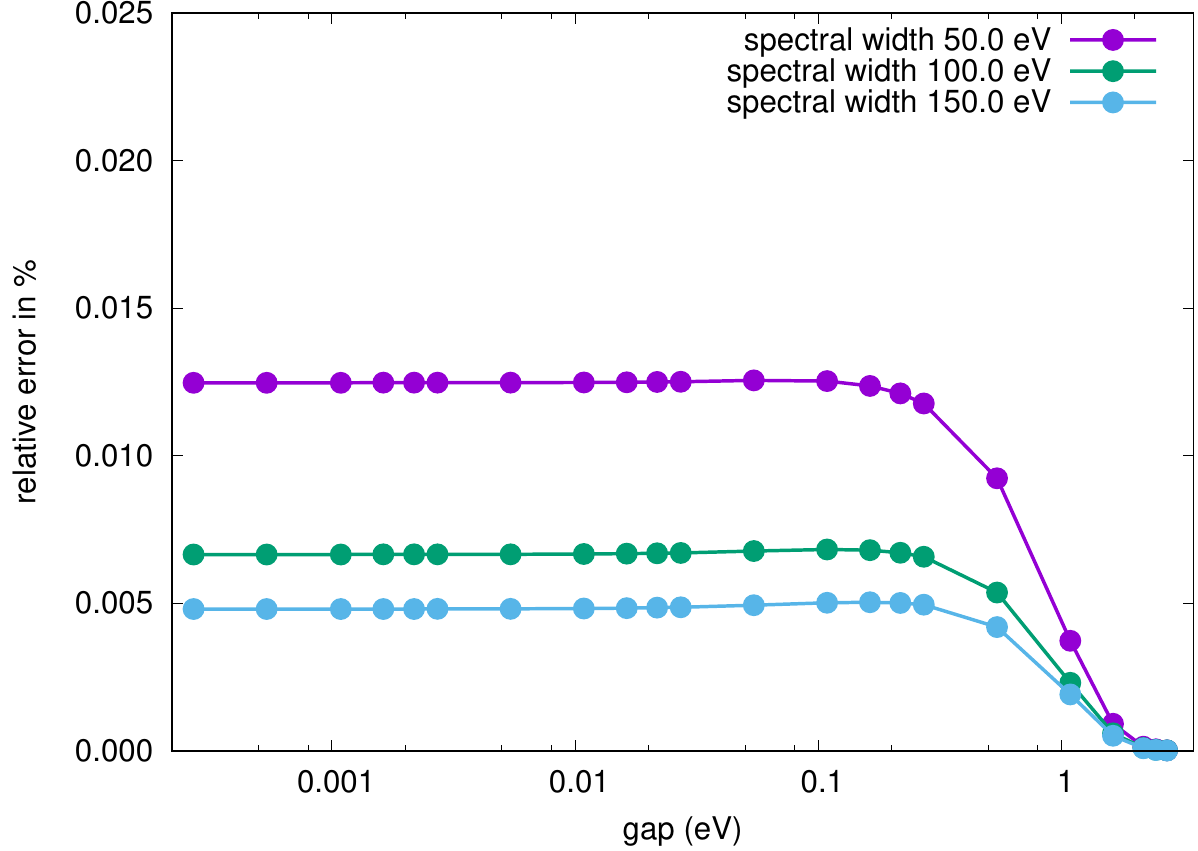}
 \caption{Difference between the energies calculated by \textsc{CheSS} using FOE and a reference LAPACK calculation, respectively, as a function of the HOMO-LUMO gap and for various spectral widths.
  The larger error for the smaller spectral widths can be explained by the eigenvalue spectrum being denser in that case, thus increasing the error introduced by the finite temperature smearing used by FOE.}
 \label{fig:accuracy_foe}
\end{figure}

\paragraph{Density kernel}
To asses the accuracy for the density kernel computation we compare the energy calculated by \textsc{CheSS} using the FOE method ($E_{FOE}=\Tr(\mathbf{K}_{FOE}\mathbf{H})$) and the one determined by a reference calculation using LAPACK (($E_{LAPACK}=\Tr(\mathbf{K}_{LAPACK}\mathbf{H})$)).
The polynomial degree was again determined automatically, leading to values between 270 and 1080.
In Fig.~\ref{fig:accuracy_foe} we show the difference between these two values as a function of the spectral width and the HOMO-LUMO gap.
As can be seen, the error shows only little variation with respect to both quantities and is always of the order of $0.01\%$.

\begin{figure}
 \includegraphics[width=1.0\columnwidth]{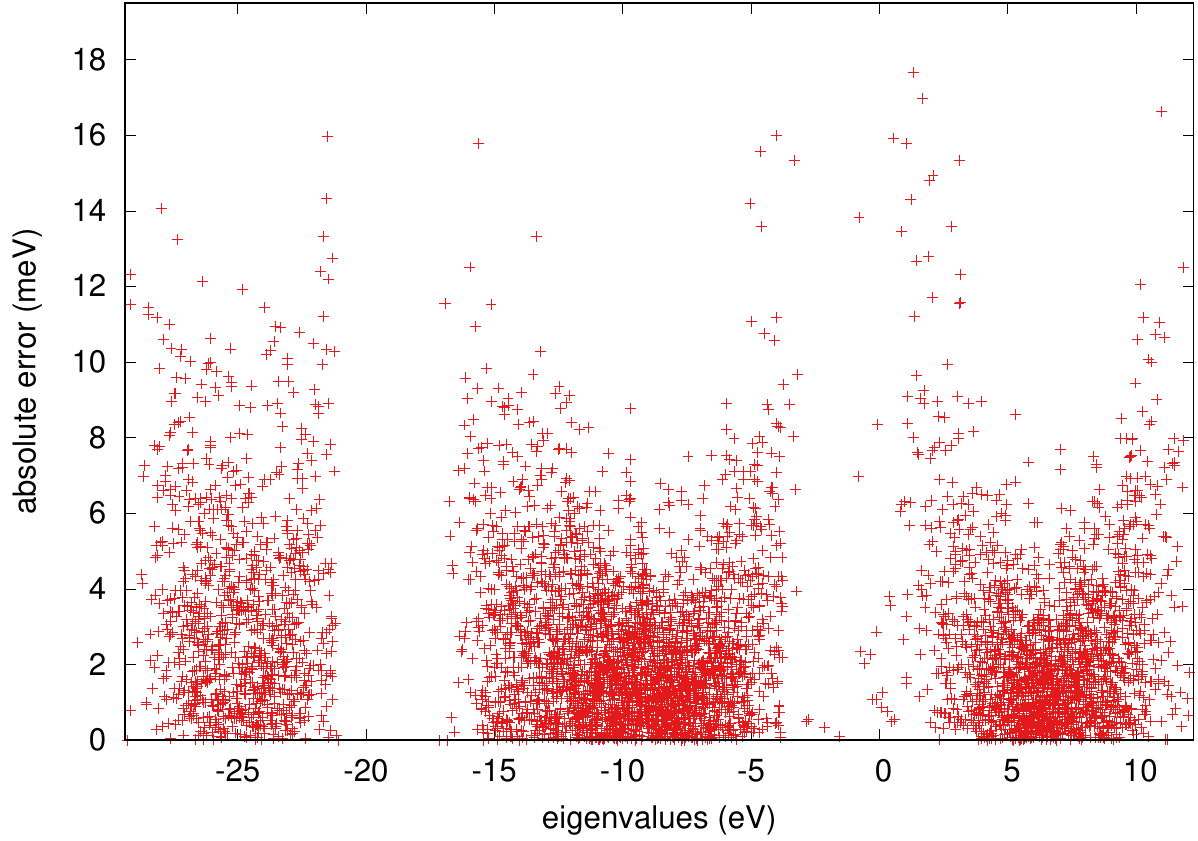}
 \caption{Errors of the calculated eigenvalues for a sparse matrix of dimension $6000 \times 6000$, with characteristics as explained in the text.
 }
 \label{fig:eigenvalues_accuracy}
\end{figure}

\paragraph{Selected eigenvalues}
For the assessment of the accuracy of the selected eigenvalues, we work directly --- i.e.\ without modifying its values --- with a matrix coming from a calculation with \textsc{BigDFT} in order to have a realistic setup.
The matrix has a spectral width of \unit[41.36]{eV}, which means that the 6000 eigenvalues only exhibit narrow separations among each other, being of the order of some meV.
The accuracy with which the eigenvalues can be calculated depends strongly on the decay length of the error function that is used;
for this test we took a value of $\beta=\unit[27.2]{meV}$, leading to a polynomial degree of 4120.
Nevertheless we see from the results in Fig.~\ref{fig:eigenvalues_accuracy} that this is enough to determine the eigenvalues quite accurately;
more precisely, the mean difference between the exact result and the one calculated by \textsc{CheSS} is only \unit[$2.6 \pm 2.3$]{meV}.

If a higher accuracy than the one obtained is required, a smaller value for $\beta$ has to be chosen.
This would, however, dramatically increase the cost due to the higher polynomial degree that is required.
Our method is thus more suited to get a rough estimate of an (arbitrary) eigenvalue rather than to calculate its exact value.
In the context of electronic structure calculations, an example for such a situation is the calculation of the HOMO-LUMO gap, where the intrinsic error of the used theory (e.g.\ DFT) is usually much larger than that of the numerical method to calculate the eigenvalues.
Additionally, in this context the polynomials of the density matrix expansion can be reused, and an estimate of the HOMO-LUMO gap can thus be obtained on the fly with hardly any extra cost, comparable to the method in Ref.~\citenum{rubensson-interior-2014}.

\subsection{Scaling with matrix properties}
\label{sec: Scaling with system properties}
In this section we want to asses the performance of \textsc{CheSS} with respect to certain specific properties of the matrices, in order to determine under which circumstances it offers the biggest benefits.
For the tests we again used the same set of matrices as in Sec.~\ref{sec: Accuracy}.
\begin{figure*}
 \subfloat[][Runtime as a function of the number of non-zero elements in the original matrix $\mathbf{M}$. The dashed lines are linear fits.]{
  \includegraphics[width=0.99\columnwidth]{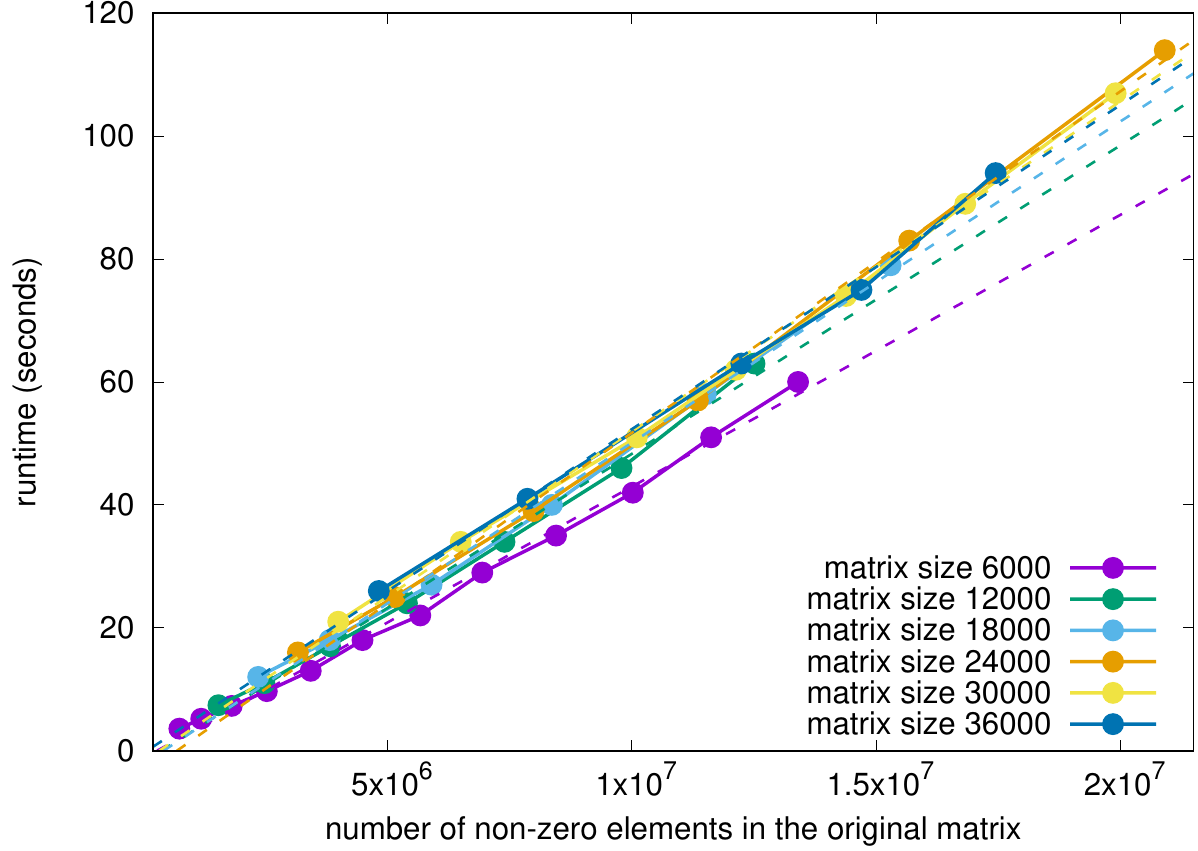}
  \label{fig:timings_size_ice_small}}
   \quad
 \subfloat[][Runtime as a function of the number of non-zero elements in the inverse $\mathbf{M}^{-1}$. The dashed lines are quadratic fits.]{
  \includegraphics[width=0.99\columnwidth]{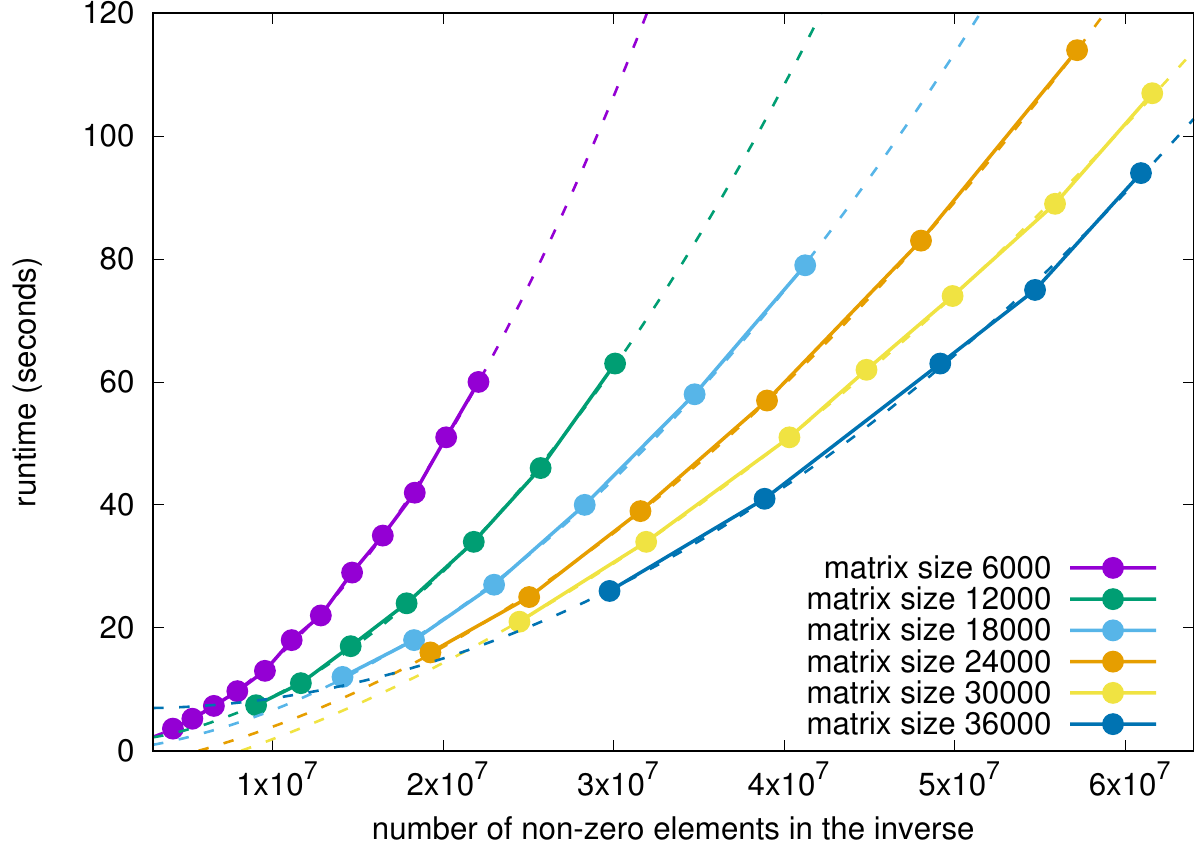}
  \label{fig:timings_size_ice_large}}
  \caption{Runtime for the calculation of the inverse as a function of the number of non-zero entries of the original matrix (Fig.~\ref{fig:timings_size_ice_small}) and the inverse (Fig.~\ref{fig:timings_size_ice_large}), for various matrix sizes and ``degrees of sparsity''.
   In Fig.~\ref{fig:timings_size_ice_small} we see a linear scaling with respect to the number of non-zero entries and hardly any dependence on the total matrix size, whereas in
   Fig.~\ref{fig:timings_size_ice_large} there is a quadratic scaling and a clear dependence 
   on the total matrix size.
   The reasons for this different behavior are discussed in the text.
   All runs were performed in parallel using 1280 cores (80 MPI tasks spanning 16 OpenMP threads each) on MareNostrum 3.
  }
 \label{fig:timings_size_ice}
\end{figure*}

\subsubsection{Scaling with matrix size and sparsity}
\label{sec: Matrix size and sparsity}

First we want to investigate how efficiently \textsc{CheSS} can exploit the sparsity of the matrices.
Since the sparsity enters via the matrix vector multiplications for the construction of the Chebyshev polynomials --- which are the same for all operations --- we only present data for the inverse.
In Fig.~\ref{fig:timings_size_ice} we show the runtime as a function of the non-zero entries of the matrix, for various matrix sizes.
Moreover we distinguish between the runtime with respect to the number of non-zero entries in the original matrix (Fig.~\ref{fig:timings_size_ice_small}) and 
the inverse (Fig.~\ref{fig:timings_size_ice_large}).
For each matrix size we generated several matrices with different ``degrees of sparsity'', i.e.\ different numbers of non-zero entries.
All matrices were prepared to have a condition number of about 11, leading to an automatically determined polynomial degree of 60, and the runs were performed in parallel using 1280 cores.

As can be seen from Fig.~\ref{fig:timings_size_ice_small}, the runtime scales linearly with respect to the number of non-zero elements in the original matrix and hardly depends on the matrix size.
This is not surprising, as the non-zero elements directly determine the cost of the matrix vector multiplications, and thus demonstrate the good exploitation of the sparsity by \textsc{CheSS}.

In Fig.~\ref{fig:timings_size_ice_large} we see, however, a non-linear behavior.
This can be explained by the fact that the cost of calculating each of the non-zero elements of the inverse depends again 
on the number of non-zero elements, yielding in total this quadratic behavior.
In addition we see here also a dependence on the total matrix size, which is due to the buffer regions.
The number of non-zero entries in the inverse, $|\mathbf{M}^{-1}|$, is related to the number of non-zero entries in the original matrix, $|\mathbf{M}|$, via $|\mathbf{M}^{-1}| = |\mathbf{M}| + c$, where $c$ depends linearly on the matrix size.
Therefore, in order to reach a given value of $|\mathbf{M}^{-1}|$, $|\mathbf{M}|$ must be larger the smaller the matrix is, which explains the higher cost for the smaller matrices.

\subsubsection{Scaling with spectral properties}
\label{sec:Scaling with spectral properties}
As mentioned earlier, the characteristics of the eigenvalue spectrum of the matrices are an important aspect for the performance of \textsc{CheSS}, and we therefore want to investigate this in more detail.

\paragraph{Condition number for the inverse}
For the calculation of the inverse we restrict ourselves to the case of positive definite matrices, which means that the spectral width can well be characterized by the condition number $\kappa$.
We prepared a set of matrices with condition numbers ranging from $6$ to $1423$, and additionally conducted calculations for two setups.
In the first one we used a default guess for the eigenvalue bounds of $[0.5,1.0]$,
whereas in the second setup we used already well adjusted values.

\begin{figure}
 \includegraphics[width=1.0\columnwidth]{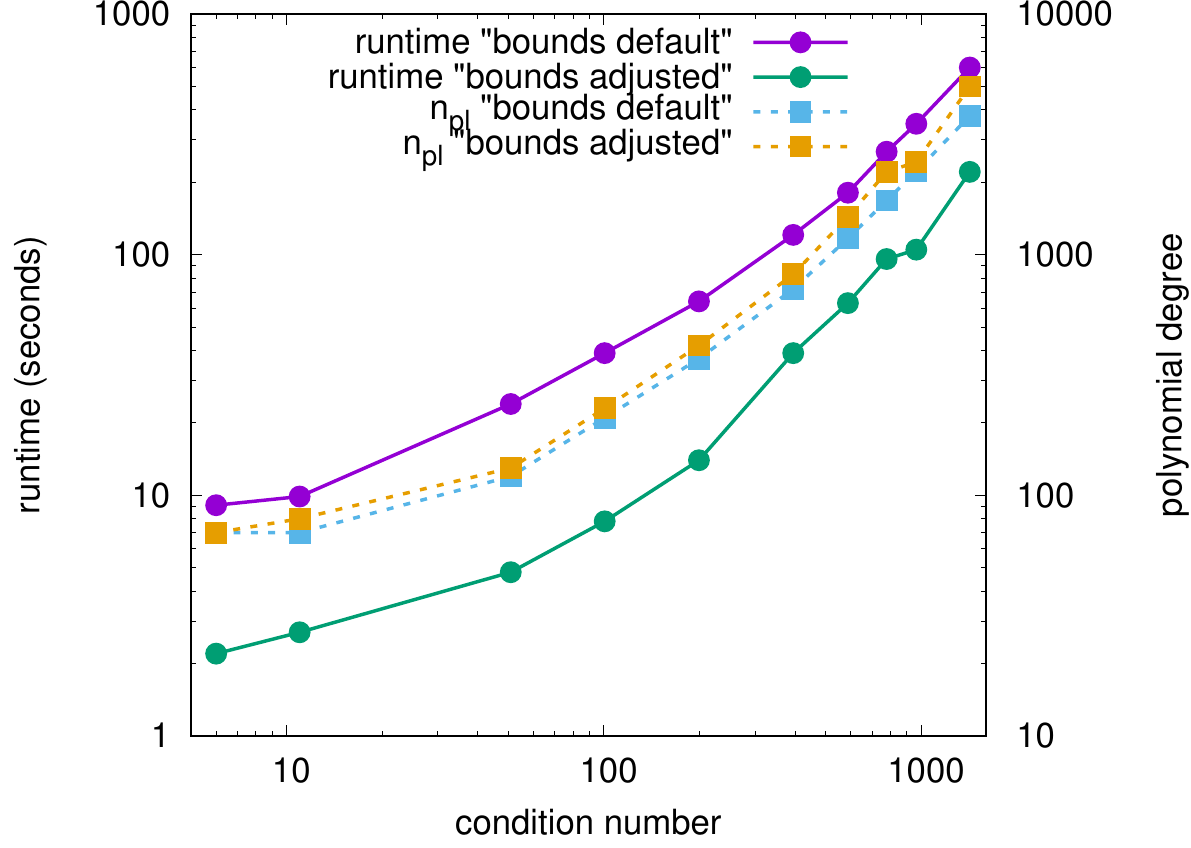}
 \caption{
 Runtime and polynomial degree $n_{pl}$ for the calculation of the inverse as a function of the condition number.
 ``bounds default'' means that the runs where started with default values for the upper and lower eigenvalue bounds, whereas ``bounds adjusted'' means that well adjusted values were used.
  All runs were performed in parallel using 480 cores (60 MPI tasks spanning 8 OpenMP threads each) on MareNostrum 3.}
 \label{fig:timings_kappa}
\end{figure}

In Fig.~\ref{fig:timings_kappa} we show the results of this benchmark, with the runs being performed in parallel using 480 cores.
As expected there is a very strong increase of the run time for larger condition numbers, due to the higher polynomial degree that is required.
However we also see that a lot can be gained by choosing a good input guess for the eigenvalue bounds --- something which is often possible in practicable applications.
In this case the polynomial degree remains more or less the same, but the expensive search for the correct eigenvalue bounds can be saved.
Moreover, looking again at the values of $\kappa$ in Tab.~\ref{tab:kappa_and_specwidth_BigDFT}, we see that the basis set employed by \textsc{BigDFT} indeed enables \textsc{CheSS} to operate in the optimal range of small condition numbers.

\begin{figure}
  \includegraphics[width=1.0\columnwidth]{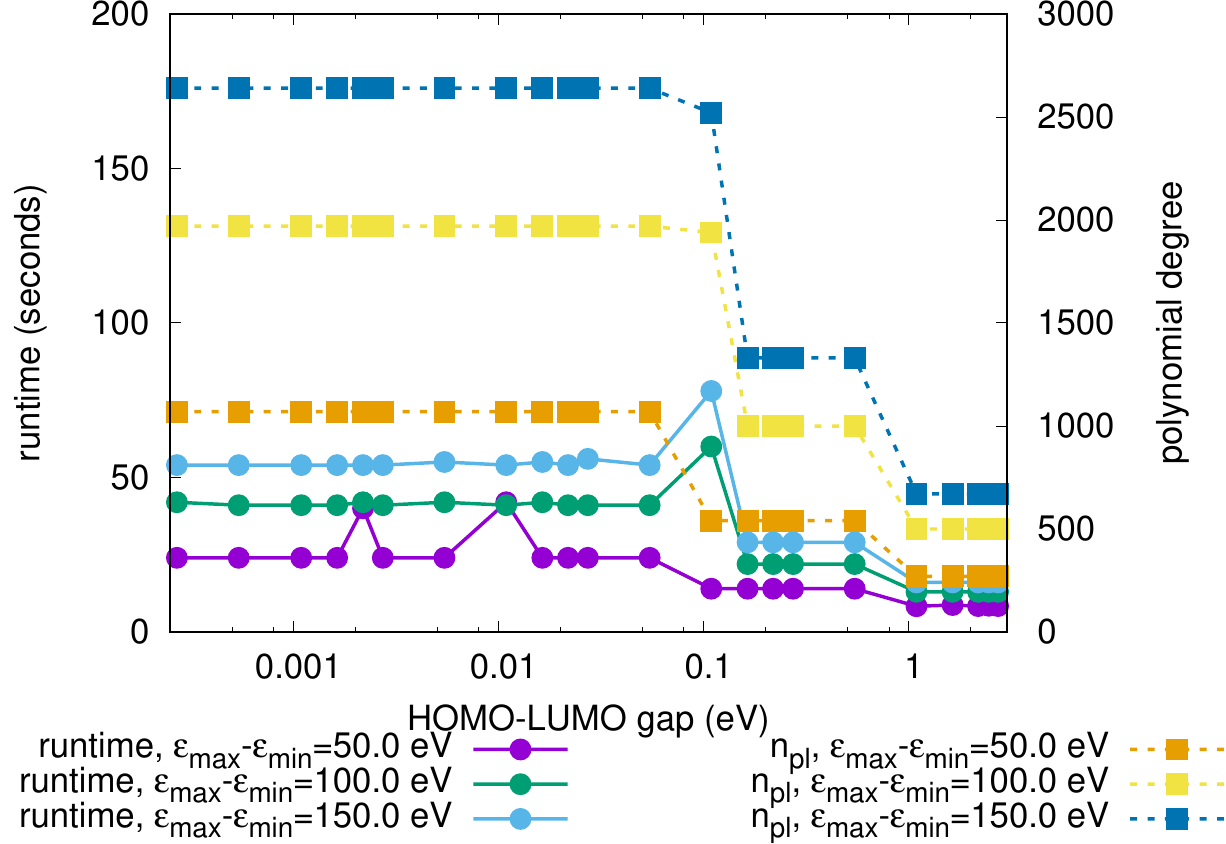}
  \label{fig:timings_foe_guessed}
 \caption{
 Runtime and polynomial degree $n_{pl}$ for the density kernel calculation as a function of the HOMO-LUMO gap, for various spectral widths. 
 The runs were started with already well adjusted values for both the upper and lower eigenvalue bounds and $\beta$,
  and performed in parallel using 480 cores (60 MPI tasks spanning 8 OpenMP threads each) on MareNostrum 3.}
 \label{fig:timings_foe}
\end{figure}

\paragraph{Spectral width and HOMO-LUMO gap for the density kernel}
For the determination of the density kernel the essential characteristic is not the condition number any more, but rather the total spectral width.
Additionally we have a dependence on the parameter $\beta$, which determines how fast the error function used to assign the occupation numbers decays between the highest occupied and the lowest unoccupied state and is therefore directly related to the HOMO-LUMO gap.
The smaller $\beta$ is, the more the error function resembles a step function, which is difficult to represent using polynomials.
As a consequence, the polynomial degree becomes very large for both large spectra and small HOMO-LUMO gaps.

In Fig.~\ref{fig:timings_foe} we show the runtime for a density kernel calculation as a function of the HOMO-LUMO gap and for various spectral widths, with the runs being performed in parallel using 480 cores.
Following the considerations of the previous test for the inverse we used already well adjusted values for the eigenvalue bounds and $\beta$.
We see our assumption confirmed, as calculations with small gaps and large spectral widths are considerably heavier.
Whereas the value of the gap is imposed by the system under investigation, the spectral width depends on the specific computational setup --- and in particular also the basis set --- that is used.
In order to keep it small, it is advisable to use a minimal basis set of optimized functions, which --- among other advantages~\cite{mohr-fragments1-2017,mohr-fragments2-2017} --- has the benefit that it only contains few virtual (and therefore high energetic) states.
These conditions are fulfilled by the basis set used by \textsc{BigDFT}, and indeed --- as shown in Tab.~\ref{tab:kappa_and_specwidth_BigDFT} --- this leads to small values for the spectral width, allowing \textsc{CheSS} to operate in an optimal range.

\subsection{Parallel scaling}
\label{sec: Parallel scaling}
The most expensive part of the \textsc{CheSS} algorithm are the matrix vector multiplications of Eq.~\eqref{eq:Chebyshev-matrix-recursion} to construct the Chebyshev matrix polynomials.
However, since this operation is independent for each vector, it can be parallelized in a straightforward way.
To account for possible non-homogeneities of the sparsity pattern we have implemented a mechanism that automatically assigns the vectors to the parallel resources such as to optimize the load balancing.
\textsc{CheSS} exhibits a two level hybrid parallelization:
On a coarser level the workload is parallelized using MPI (i.e.\ distributed memory parallelization), whereas on a finer level an additional parallelization using OpenMP (i.e.\ shared memory parallelization) is used.
In this way it is possible to obtain a very efficient exploitation of parallel resources.

\begin{figure*}
 \subfloat[][Speedup with respect to the minimal number of cores that was possible due to memory restrictions (80, 160 and 320, respectively). To ease the comparison, the curves for the 24000 matrix and 36000 matrix start at 2.0 and 4.0, respectively.]{
  \includegraphics[width=0.99\columnwidth]{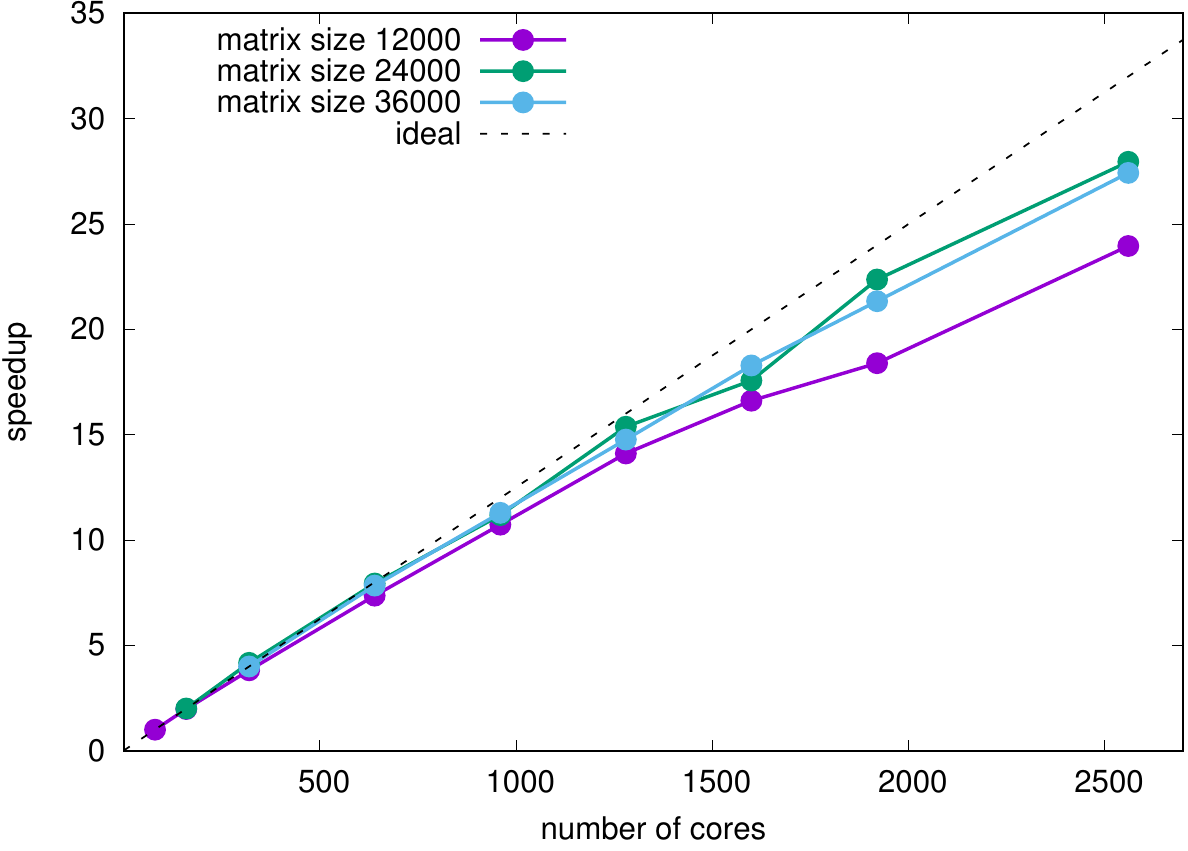}
  \label{fig:timings_procs_ice_speedup}}
  \quad
 \subfloat[][Absolute runtime for all runs.]{
  \includegraphics[width=0.99\columnwidth]{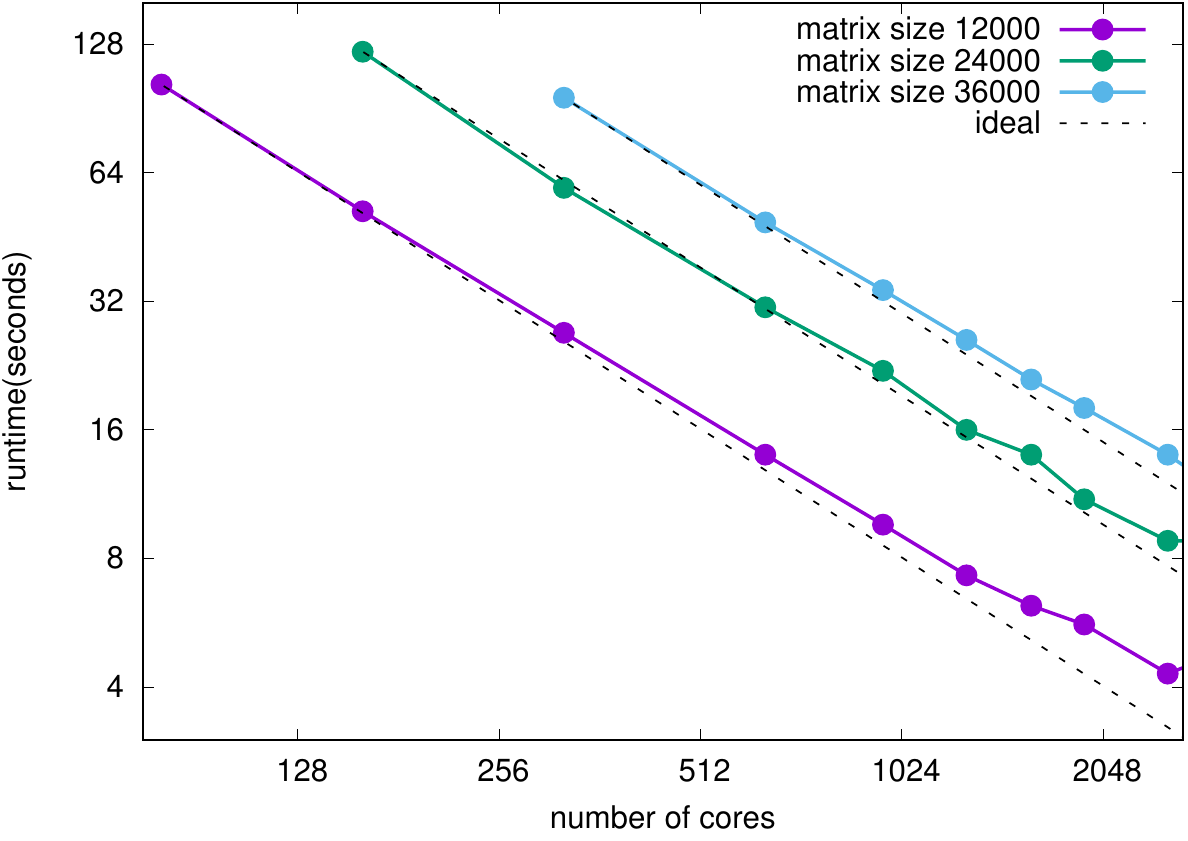}
  \label{fig:timings_procs_ice_time}}
 \caption{Parallel scaling for the calculation of the inverse using \textsc{CheSS}. Fig.~\ref{fig:timings_procs_ice_speedup} shows the speedup, whereas Fig.~\ref{fig:timings_procs_ice_time} shows the absolute runtimes. All the runs where performed with 16 OpenMP threads, and only the number of MPI tasks was varied. The benchmarks were done on MareNostrum 3.}
 \label{fig:timings_procs_ice}
\end{figure*}

\paragraph{Scaling for small to medium size matrices}
In Fig.~\ref{fig:timings_procs_ice} we show the parallel scaling for the calculation of the inverse; for the sake of simplicity we again focused on this case, as the results for the other operations would be very similar.
We took matrices of three different sizes ($12000 \times 12000$, $24000 \times 24000$ and $36000 \times 36000$), with the number of nonzero entries chosen to be approximately proportional to the matrix size, 
and varied the number of cores from --- depending on the matrix size due to memory limitations --- 80, 160 and 320, respectively, up to 2560.
The condition number for all matrices was set to 11, which led to an (automatically determined) polynomial degree of 60.

In Fig.~\ref{fig:timings_procs_ice_speedup} we show the speedup with respect to the minimal number of cores.
The curves are very similar for all matrices, meaning that a good exploitation of parallel resources --- and consequently speedup ---
can already be obtained for small systems.
Indeed it is possible to bring down the calculation time to only a few seconds if enough computational resources are available, as can be seen from Fig.~\ref{fig:timings_procs_ice_time}.
By fitting the data for the $6000 \times 6000$ matrix --- which shows the worst scaling --- to Amdahl's law~\cite{amdahl-validity-2007}, we get an overall (i.e.\ also including communications) parallel fraction of 98.9\%, which gives a maximal theoretical speedup of about 90.
However, it must be stressed that this gives the speedup with respect to 80 cores, and hence the overall maximal speedup is considerably higher.

\begin{figure}
 \includegraphics[width=1.0\columnwidth]{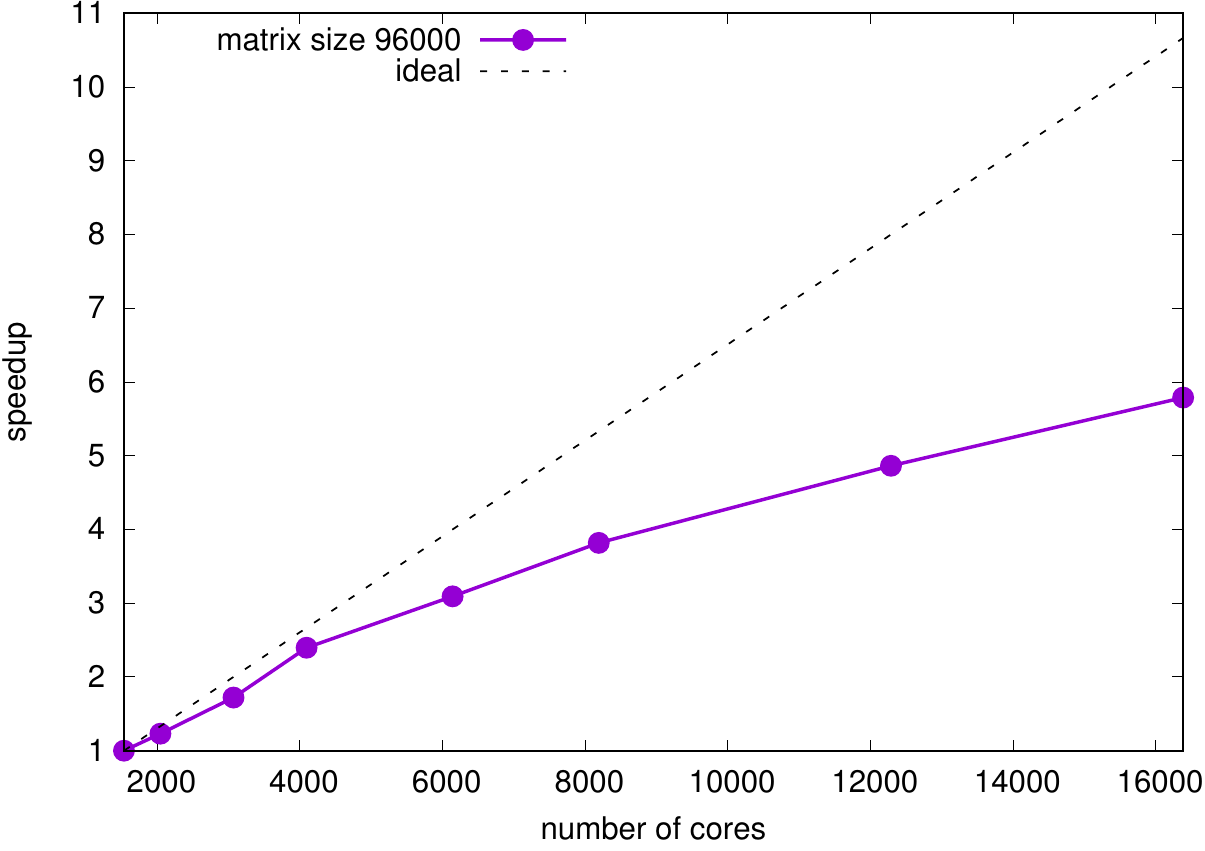}
 \caption{
  Extreme scaling behavior of \textsc{CheSS} for the calculation of the inverse, going from 1536 cores up to 16384 cores.
  The runs were performed using 8 OpenMP threads, only varying the number of MPI tasks, and the speedup is shown with respect to the first data point.
  The benchmarks were done on the K computer.
  }
 \label{fig:ice_extreme_scaling}
\end{figure}

\paragraph{Extreme scaling behavior}
In the previous paragraph we have demonstrated the efficient exploitation of the typically used parallel resources for small to medium size systems.
Now we also want to show the extreme scaling behavior of \textsc{CheSS}, i.e.\ how the library behaves when going to ten thousands of cores.
For this purpose we chose a slightly larger matrix, namely of size $96000 \times 96000$, again stemming from a calculation of a water droplet with \textsc{BigDFT}.
In Fig.\ref{fig:ice_extreme_scaling} we show the scaling that we obtain for the calculation of the inverse of this matrix, going from 1536 cores up to 16384 cores.
Considering that the chosen matrix is still not extremely big, \textsc{CheSS} scales reasonably well also for very large number of cores, demonstrating its capability to perform efficient calculations under extremely parallel conditions.

\subsection{Comparison with other methods}
\label{sec: Comparison with other methods}
Finally we want to compare the performance of \textsc{CheSS} with two other methods that allow to perform the same operations.
On the one hand we benchmark it against \textsc{(Sca)LAPACK}, which is presumably the most efficient way to perform general purpose linear algebra operations for dense matrices.
On the other hand we compare it with \textsc{PEXSI}, which can exploit the sparsity of the matrices and is an established package 
for large scale DFT calculations,
as demonstrated for instance by its coupling with the \textsc{SIESTA} code~\cite{lin-SIESTA-PEXSI-2014}.

We tested five sets of matrices, ranging from $6000 \times 6000$ to $30000 \times 30000$, with the number of non-zero elements proportional to the matrix size.
Moreover we performed the comparison for various values of the spectral width, in order to assess this important dependence.
Following the conclusions of Sec.~\ref{sec:Scaling with spectral properties} we started the \textsc{CheSS} runs with well adjusted guesses for the eigenvalue bounds, thus simulating the conditions in a real application.
We performed all runs in parallel, using 160 MPI tasks with each one spanning 12 OpenMP threads, i.e.\ using in total 1920 cores.
We note that such a high number of threads does not seem to be optimal for \textsc{PEXSI} due to only moderate OpenMP speedup.
This is in contrast to \textsc{LAPACK} and \textsc{CheSS}, which can exploit this wide shared memory parallelism in a very efficient way.
However such a setting might likely be imposed by the application using the library
--- i.e.\ the electronic structure code, in our case \textsc{BigDFT} ---, for instance due to memory restrictions,
or the usage of many-core systems --- becoming more and more abundant --- that are designed for shared memory parallelization,
and overall our setup thus corresponds to realistic situations.
In other terms, we are not just comparing the various solvers, but rather the solvers within a given specific (but nevertheless realistic) setup.
Nevertheless, we will show for completeness also some results using exclusively an MPI parallelization, in order to see the effects on both \textsc{CheSS} and \textsc{PEXSI}

\paragraph{Inverse}
Due to the general character of \textsc{(Sca)LAPACK}, we have to implement the required functionality on our own.
To calculate matrix powers $\mathbf{M}^a$, we first diagonalize the matrix $\mathbf{M}$ as $\mathbf{D}=\mathbf{U}^{T}\mathbf{M}\mathbf{U}$, with a diagonal matrix $\mathbf{D}$ and a unitary matrix $\mathbf{U}$.
Then we can easily apply the desired power to the diagonal elements $D_{ii}$ in order to get $\mathbf{D}^a$,
and finally we obtain the desired result as $\mathbf{M}^a=\mathbf{U}\mathbf{D}^{a}\mathbf{U}^{T}$.
The diagonalization was done using the PDSYEVD routine, which is based on a parallel divide-and-conquer algorithm.
For the calculation of the inverse,
there exist more specific routines within \textsc{(Sca)LAPACK}; however we nevertheless use the aforementioned approach 
as it is the most general one and allows --- as does \textsc{CheSS} --- to calculate any desired power.
With respect to PEXSI, we can invert a matrix using the Selected Inversion algorithm, which this package contains as well as it uses it for the pole expansion within the density kernel calculation.
\begin{figure*}
 \includegraphics[width=1.0\textwidth]{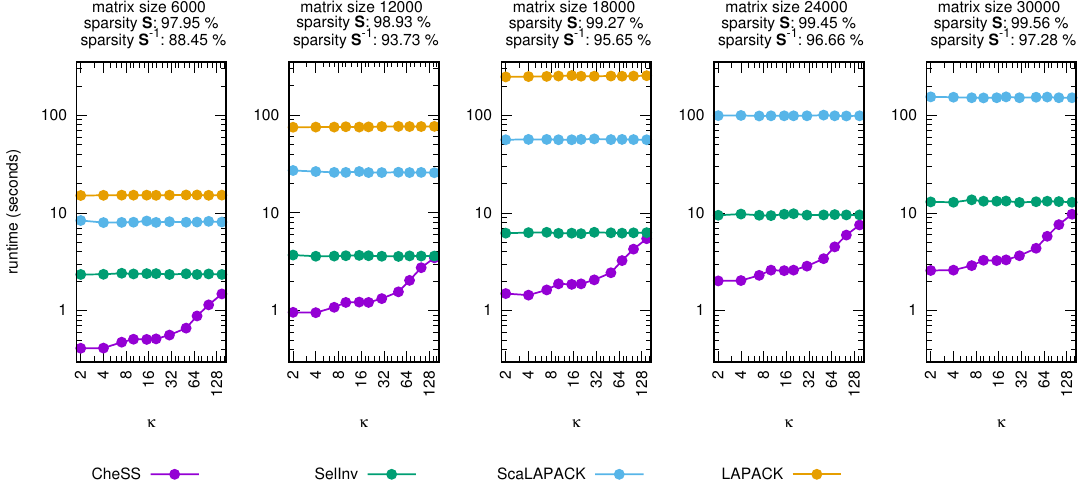}
 \caption{Comparison of the runtimes for the matrix inversion using \textsc{CheSS}, the Selected Inversion from \textsc{PEXSI}, \textsc{ScaLAPACK} and \textsc{LAPACK}, for various matrices and as a function of the condition number.
  All runs were performed in parallel, using 1920 cores (160 MPI tasks spanning 12 OpenMP threads each).
  The \textsc{CheSS} runs were started with well adjusted bounds for the eigenvalue spectrum, and the polynomial degree ranged from 60 to 260.
  For \textsc{LAPACK}, no results for matrices larger than 18000 are shown due to their long runtime.
  The benchmarks were done on MareNostrum 4.
  }
 \label{fig:timings_comparison_ice}
\end{figure*}

In Fig.~\ref{fig:timings_comparison_ice} we show the timings that we obtain for the calculation of the inverse as a function of the condition number $\kappa$.
As can be seen, \textsc{CheSS} is indeed the most efficient method for matrices with small values of $\kappa$.
The only method that is competitive is the Selected Inversion, which does not show any dependence on the condition number and will therefore be faster for large values.
The crossover between \textsc{CheSS} and the Selected Inversion depends on the matrix size --- thanks to the linear scaling property of \textsc{CheSS} it is higher the larger the matrices are.
Nevertheless we see that for all matrices used in this test \textsc{CheSS} is the most efficient method, with the crossover with respect to the condition number being located at about 150.
Following the discussion in Sec.~\ref{sec: Applicability of CheSS for electronic structure calculation} this is a value that is easily reachable in practical applications.
Last but not least, we note that there is no case where \textsc{LAPACK} or \textsc{ScaLAPACK} are the fastest methods, demonstrating the need to exploit the sparsity of the matrices.

\paragraph{Density kernel calculation}
Since \textsc{CheSS} is mainly designed for systems with a decent HOMO-LUMO gap, we are focusing on such systems;
more specifically we always set the gap to a value of \unit[1]{eV}.
\begin{figure*}
 \includegraphics[width=1.0\textwidth]{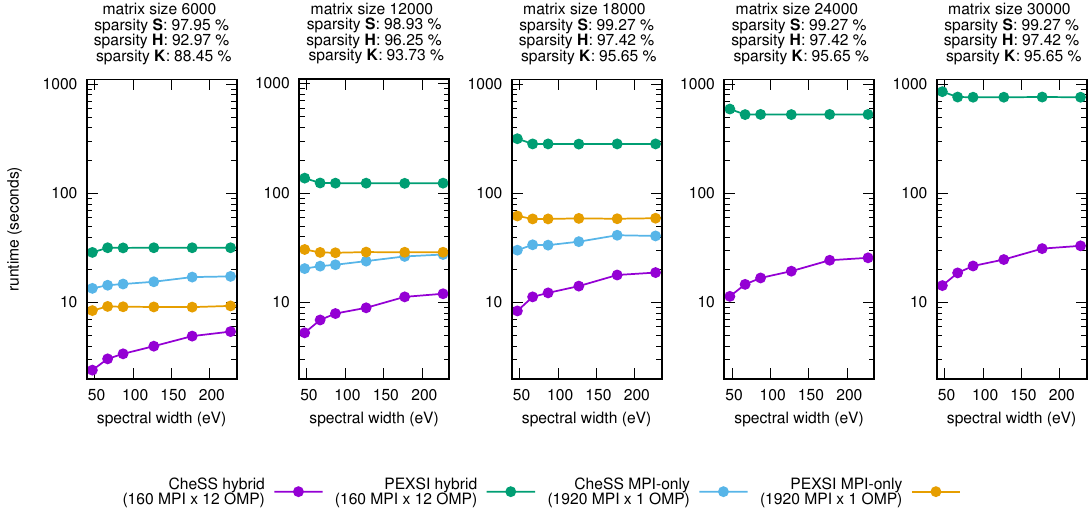}
 \caption{Comparison of the runtimes for the density kernel calculation using \textsc{CheSS} and \textsc{PEXSI},
 for various matrices and as a function of the spectral width.
  All matrices had a HOMO-LUMO gap of \unit[1]{eV}.
  The runs were performed in parallel using 1920 cores,
  once with a hybrid setup (160 MPI tasks spanning 12 OpenMP threads each) and once with an MPI-only setup.
  The \textsc{CheSS} runs were started with a good input guess for the eigenvalue bounds, with the polynomial degree ranging from 270 to 790.
  For \textsc{PEXSI}, the number of poles used for the expansion was set to 40, following Ref.~\citenum{lin-2013-accelerating}.
  The benchmarks were done on MareNostrum 4.
  }
 \label{fig:timings_comparison_foe_withMPIonly}
\end{figure*}
In Fig.~\ref{fig:timings_comparison_foe_withMPIonly} we show the runtimes that we obtain for \textsc{CheSS} and \textsc{PEXSI}
as a function of the spectral width.
Moreover we show results for both a hybrid setup (160 MPI tasks spanning 12 OpenMP threads each) and an MPI-only setup (1920 MPI tasks).
We do not show a comparison with \textsc{(Sca)LAPACK}, 
since the tests for the inverse have demonstrated the need of using methods that take into account the sparsity of the matrices.

When using the hybrid setup,
\textsc{CheSS} is in all cases the most efficient approach even though the runtime slightly increases as a function of the spectral width.
\textsc{PEXSI} does not exhibit such a a dependence, but the difference to \textsc{CheSS} is so large that --- according to the spectral widths presented in Sec.~\ref{sec: Applicability of CheSS for electronic structure calculation} --- it is easily possible to always operate in the regime where \text{CheSS} is the most efficient approach.
When using the MPI-only setup, the runtimes of \textsc{CheSS} systematically worsen, whereas those of \textsc{PEXSI} improve.
This leads to an inversion of the ranking for the smallest matrix, but for all the other ones \textsc{CheSS} remains the fastest method.
Finally it is important to mention that the MPI-only setup does not allow calculations beyond matrix sizes of 18000 due to memory limitations.
This clearly demonstrates the need of performing calculations using a hybrid distributed memory / shared memory approach --- a regime in which \textsc{CheSS} is clearly superior.

\section{Conclusions and outlook}
\label{sec: Conclusions and outlook}
We presented \textsc{CheSS}, the ``Chebyshev Sparse Solvers'' library, which implements the flexible and efficient computation of matrix functions using an expansion in Chebyshev polynomials.
The library was developed in the context of electronic structure calculations --- in particular DFT --- with a localized basis set,
but can also be extended and applied to other problems.
More specifically, \textsc{CheSS} can calculate the density matrix, any --- in particular also non-integer --- power of a matrix, and selected eigenvalues.
\textsc{CheSS} is capable to efficiently exploit the sparsity of the matrices, scaling linearly with the number of non-zero elements, and is consequently well suited for large scale applications requiring a linear scaling approach.

The performance of \textsc{CheSS} for a specific problem depends on how well the matrix function can be approximated by Chebyshev polynomials.
This depends, obviously, on the function itself, but also on the spectral width of the matrices.
Whereas the first dependence is imposed by the specific application, the second one is as well related to the physical model and basis set that is employed.
\textsc{CheSS} has been designed for matrices exhibiting a small eigenvalue spectrum, since this reduces the number of polynomials required to represent the function that shall be calculated.
We used the library together with the DFT code \textsc{BigDFT}, which uses a \emph{minimal} set of \emph{quasi-orthogonal in-situ optimized} basis functions, leading to the required small eigenvalue spectra of the matrices.
We showed that in such a favorable setup \textsc{CheSS} is able to clearly outperform other comparable approaches, and hence can considerably boost large scale DFT calculations.

Finally, the algorithm on which \textsc{CheSS} is built can be parallelized in a very efficient way, allowing the library to scale up to thousands of cores.
In addition, the parallelism can already be well exploited for relatively small matrices, and consequently good speedups and low runtimes can be obtained for such systems.
The initial performance of \textsc{CheSS} was evaluated within an performance audit by the Performance Optimization and Productivity center of excellence (POP), which helped to understand performance issues and gave recommendations and performance improvements~\cite{POP-audit}.
In addition, we continue to cooperate with POP
to further analyze and optimize the parallel efficiency and scalability of the library. 

\textsc{CheSS} is been used intensively within the BigDFT code~\cite{mohr-2014-daubechies,mohr-2015-accurate} and is about to being coupled with the SIESTA code~\cite{soler-the_siesta-2002,artacho-the_siesta-2008}.
Moreover it should be possible for any code working with localized basis functions to use this library and hence to accelerate large scale calculations.

\section{Acknowledgments}
We gratefully acknowledge the support of the MaX (SM) and POP (MW) projects, 
which have received funding from the European Union's Horizon 2020 research and innovation programme
under grant agreement No. 676598 and 676553, respectively.
This work was also supported by the Energy oriented Centre of Excellence
(EoCoE), grant agreement number 676629, funded within the Horizon2020
framework of the European Union, as well as by the Next-Generation Supercomputer project (the K computer project) and the FLAGSHIP2020 within the priority study5 
(Development of new fundamental technologies for high-efficiency energy creation, conversion/storage and use) from the Ministry of Education, Culture, Sports, 
Science and Technology (MEXT) of Japan.
We (LG, DC, WD, TN) gratefully acknowledge the joint CEA-RIKEN collaboration action.

\bibliography{citationlist}

%

\end{document}